\documentclass{jfm}

\usepackage{graphicx}
\usepackage{newtxtext}
\usepackage{newtxmath}
\usepackage{natbib}
\usepackage{hyperref}
\usepackage{amssymb}
\hypersetup{
    colorlinks = true,
    urlcolor   = blue,
    citecolor  = black,
}

\newcommand{\RomanNumeralCaps}[1]
\linenumbers

\usepackage{tikz}
\usepackage{pgfplotstable}
\usetikzlibrary{hobby,decorations.pathreplacing,calligraphy}

\graphicspath{{./figs/}}

\title{Diagnosing tracer transport in convective penetration of a stably stratified layer}

\author{Charles W. Powell\aff{1}
	\corresp{\email{cwp29@cam.ac.uk}},
	Peter H. Haynes\aff{1}
	\and John R. Taylor\aff{1}}

\affiliation{\aff{1}DAMTP, University of Cambridge, Centre for Mathematical Sciences, Wilberforce Road,
	Cambridge CB3 0WA, UK}

\begin{document}
\maketitle

% Shorthand commands for simulation data
\newcommand{\bflux}{3.96\times10^{-7}}
\newcommand{\zmaxval}{3.94}
\newcommand{\zsourceval}{-7.97}
\newcommand{\tmaxval}{6.75}

% Commands for convenience
\newcommand{\Reb}{\Rey_b}
\renewcommand{\Pr}{Pr}
\newcommand{\benv}{b^{(\text{env})}}
\newcommand{\zmax}{z_{\max}}
\newcommand{\kappaSGS}{\kappa_{\text{SGS}}}
\newcommand{\nuSGS}{\nu_{\text{SGS}}}
\newcommand{\nutot}{\nu_{\mathrm{tot}}}
\newcommand{\kappatot}{\kappa_{\mathrm{tot}}}

\begin{abstract}
    We use large-eddy simulations to study the penetration of a buoyant plume carrying a passive
    tracer into a stably stratified layer with constant buoyancy frequency. Using a buoyancy-tracer
    volume distribution, we develop a method for objectively partitioning plume fluid in buoyancy-tracer 
    space into
    three regions, each of which corresponds to a coherent region in physical space. Specifically,
    we identify a source region where undiluted plume fluid enters the stratified layer, a transport
    region where much of the transition from undiluted to mixed fluid occurs in the plume cap, and
    an accumulation region corresponding to a radially spreading intrusion. This method enables
    quantification of different measures of turbulence and mixing within each of the three regions,
    including potential energy and turbulent kinetic energy dissipation rates, an activity parameter,
    and the instantaneous mixing efficiency. We find that the most intense buoyancy gradients lie
    in a thin layer at the cap of the penetrating plume. This provides the primary stage of mixing between 
    plume and environment and exhibits a mixing efficiency around 50\%. Newly generated mixtures of 
    environmental and plume fluid join the intrusion and experience relatively weak turbulence and 
    buoyancy gradients. As the intrusion spreads radially, environmental 
    fluid surrounding the intrusion is mixed into the intrusion with moderate mixing efficiency. This
    dominates the volume of environmental fluid entrained into the region containing plume fluid. However, the
    `strongest' entrainment, as measured by the specific entrainment rate, is largest in the plume cap
    where the most buoyant environmental fluid is entrained.
\end{abstract}


%%%%%%%%%%%%%%%%%%%%%%%%%%%%%%%%%%%%%%%%%%%%%%%%%%%%%%%%%%%%%%%%%%%%%%%%%%%%%%%%%%%%%%%%%%%%%%%%%%%%%%%%%%%%%

\section{Introduction}

The interaction between active convection and neighbouring stably stratified regions is relevant to many
geophysical flows. An important example is the tropical upper troposphere and lower stratosphere, where
convective plumes generated by strong thunderstorm complexes can penetrate through the tropical tropopause
layer into the lower stratosphere, resulting in vertical transport of trace gases and 
water vapour \citep{randel2013, jensen2007}. Numerical simulations of convective penetration events have been performed 
using realistic and complex meteorological models \citep{dauhut2015, dauhut2018} containing many physical 
processes but these are computationally expensive and challenging to interpret. 

Another geophysical process where this fluid dynamical problem is relevant is deep convection in the open ocean. 
Typically, mixing between the deep ocean and near-surface water is hindered by the strong vertical density gradients of 
the thermocline. In some regions, including several locations in high latitude oceans and the Mediterranean Sea, intense 
buoyancy loss from the ocean surface to the atmosphere results in strong, deep-reaching convection 
\citep{marshall1999,herrmann2008}. The transport of surface water into the deep ocean sets and maintains the properties 
of the abyss \citep{marshall1999}, both in terms of the general circulation and also biogeochemical cycles 
\citep{ulses2021}. 

Further examples of naturally occurring flows
involving penetrative convection include modification of downslope oceanic gravity currents by near-surface
convection \citep{doda2023}, smaller-scale atmospheric convection below an inversion \citep{kurbatskii2001}, 
volcanic eruptions that penetrate into the stratosphere (e.g. \citet{carazzo2008, textor2003}), 
as well as the internal structure of many stars where a convectively unstable layer is bounded above and below
by stable layers \citep{singh1994, masada2013}. Moreover, the fluid dynamical problem of convective
penetration itself is of continuing scientific interest, in particular the generation of gravity waves by the
penetrating plume cap \citep{ansong2010} and the energetics of the system \citep{chen2023}, with applications
to many problems -- see \citet{hunt2015} for a discussion of open questions on fountains, i.e. plumes surrounded by a 
more buoyant environment. Much previous work has focused on laboratory studies of buoyant plumes with simple 
background density profiles, but numerical simulation has recently become feasible \citep{alfonsi2011}.

Progress towards understanding the 
contribution of convective penetration to tracer transport in geophysical settings can be made by considering an 
idealised representation of the problem in which a region of strong stable stratification is penetrated by a turbulent 
buoyant plume generated in a region with weak or zero stratification. The objective of the study reported in this paper 
is to diagnose the irreversible diapycnal tracer transport that results from turbulent 
mixing between plume fluid carrying a passive tracer and the surrounding environmental fluid in the stratified layer. 
We aim to provide a quantitative description of the mixing involved in this diapycnal transport. Such 
descriptions are essential in forming parameterisations of convective penetration. Throughout the flow evolution, 
plume fluid may be distinguished from environmental fluid by the presence of non-zero tracer concentration. Crucially, 
both the tracer concentration and buoyancy fields are subjected to turbulent mixing, resulting in the entrainment of 
environmental fluid into the plume and modification of the relationship between buoyancy and tracer within the plume. 
\citet{plumb2007} introduced a tracer-tracer probability density function to study rapid isentropic mixing in the 
stratosphere. \citet{penney2020} utilised this method to study diapycnal mixing
of passive tracers by Kelvin-Helmholtz billows arising in a stratified shear flow. Using buoyancy as one of the 
tracers, the redistribution of fluid in buoyancy-tracer space was used to interpret the mixing process. 

In this paper 
we use a similar formulation to diagnose the diapycnal transport of a passive tracer in a buoyant plume penetrating a 
linearly stably stratified layer. The numerical method is detailed in section~2. The evolution of the flow and tracer 
concentration is presented in section~3. In section~4 we introduce our formulation of the buoyancy-tracer `volume 
distribution'. We show that the flow can be partitioned into three regions of 
buoyancy-tracer space: the source region where plume fluid enters the stratified layer, a transport region through which volume 
flows during initial mixing between the plume and environment, and an accumulation region where mixed fluid settles 
and homogenises. Each of these regions of buoyancy-tracer space correspond to coherent regions of physical space that 
identify the essential structures of the flow, namely the rising plume, plume cap, and radially-spreading intrusion, 
respectively. These structures are indicated in figure~\ref{fig:schematic}. In section~5 we analyse diagnostics of the 
mixing process in each of these regions.

%%%%%%%%%%%%%%%%%%%%%%%%%%%%%%%%%%%%%%%%%%%%%%%%%%%%%%%%%%%%%%%%%%%%%%%%%%%%%%%%%%%%%%%%%%%%%%%%%%%%%%%%%%%%%

\section{Governing equations and numerical model}
\label{sec:sim}
\begin{figure}
	\centering
	\includegraphics[width=.9\textwidth]{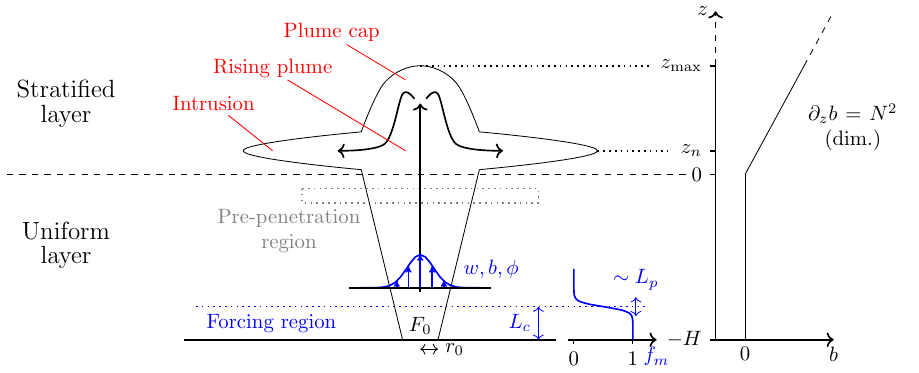}
	\caption{Setup for numerical simulations of a buoyant plume with source integral buoyancy and tracer flux $F_0$ and 
            source radius $r_0$ penetrating from a uniform layer into a linearly stably stratified layer with constant buoyancy frequency $N$. The main plume structures discussed throughout this paper are indicated in red. The pre-penetration region defined in \S\ref{sec:flow} for diagnostic purposes is shown in gray. The initial buoyancy profile (right) in the stratified environment is $b(\boldsymbol{x},0) = N^2 z$ (dimensional) for $z \ge 0$. The maximum penetration height $\zmax$ and intrusion height $z_n$ above the bottom of the stratified layer are indicated by dotted lines. We also show the forcing region of depth $L_c$ and the forcing modulation profile $f_m(z)$ decaying over a distance $L_p$ in blue, detailed in appendix~\ref{app:A}. The (azimuthally averaged) Gaussian profiles of $w, b$ and $\phi$ in the plume rising through the uniform layer are also illustrated in blue. }
	\label{fig:schematic}
\end{figure}

We consider the penetration of a buoyant plume with source radius $r_0$ and source integral buoyancy flux
$F_0$ generated in a uniform layer of depth $H$ into a stably stratified layer with buoyancy frequency $N$.
The problem setup is shown in figure~\ref{fig:schematic}. To aid in the examination of the flow evolution and
mixing, we include a passive tracer $\phi$ that satisfies the same evolution equation as buoyancy $b =
-g\rho'/\rho_0$ where $\rho'$ is the density deviation from a reference value $\rho_0$. The tracer is passive
in the sense that it has no coupling with the momentum equation. The scalar
field $\phi(\boldsymbol{x}, t)$ represents the (dimensionless) tracer concentration normalised by the tracer 
concentration on the plume centreline at the source. As illustrated in figure~\ref{fig:schematic}, we define the 
bottom of the initial stratified layer to be $z=0$. We also define $t=0$ as the time at which the plume first penetrates 
into the stratified layer. The plume source (at the base of the domain) lies at $z=-H \approx \zsourceval$ for the 
parameter choices given in table~\ref{tab:params}. The initial conditions are $\phi(\boldsymbol{x},0) = 
0$ throughout the domain whilst $b(\boldsymbol{x},0) = 0$ in the uniform layer $z \le 0$ and $b(\boldsymbol{x},0) = 
N^2 z$ (dimensional) in the stratified layer $z \ge 0$. 

We generate the buoyant plume by forcing the vertical velocity $w$, buoyancy $b$ and tracer concentration $\phi$ in 
the shallow forcing region of depth $L_c$ indicated in figure~\ref{fig:schematic}. The plume centreline lies at the 
middle of the computational domain, $x=y=0$. Details on the plume forcing method can be found in appendix~\ref{app:A}. 
At the source, the generated plume has integral buoyancy flux $F_0 = 2\int_0^\infty \left. 
\overline{w}\overline{b}\right|_{z=-H}\,r\mathrm{d}r$, where $\overline{\cdot}$ denotes an azimuthal and time average, with 
no excess momentum flux (i.e. a `pure' plume, see appendix~\ref{app:A} for detail). Turbulence is initiated in the plume by applying a 10\% perturbation to the 
forcing profiles in the forcing region and to all velocity components in the two grid layers above the forcing 
region. Turbulence develops as the plume rises through the uniform layer and we ensure that, prior to penetrating the 
stratified layer, the azimuthally averaged vertical velocity, buoyancy and tracer concentration are self-similar with 
a Gaussian radial profile as expected in a fully developed plume (see appendices~\ref{app:A} and~\ref{app:B}).

We non-dimensionalise using the source integral buoyancy flux $F_0$ (with dimensions $L^4 T^{-3}$) and
buoyancy frequency $N$ in the stable layer. The length scale is $L = F_0^{1/4} N^{-3/4}$ and the time 
scale is $T = N^{-1}$. We assume the velocity scale is $L/T$. The length scale $L$ naturally arises 
from the \citet{mtt} plume equations in a stably stratified environment. Following previous 
experimental and numerical studies (e.g \citet{briggs1965,devenish2010}), both the maximum height of the plume 
$\zmax$ and the height of the intrusion $z_n$ above the base of the stratification (illustrated in  
figure~\ref{fig:schematic}) scale with $L$ (in the case of a `lazy' plume, when incident momentum is negligible, as 
considered here).  

Owing to the large range of length scales involved in convective penetration, resolving turbulent scales with direct 
numerical simulation is not feasible due to the computational cost of very high resolution simulations.  We therefore 
use large eddy simulation (LES) with the anisotropic minimum dissipation (AMD) eddy-viscosity model to represent 
unresolved scales \citep{taylor2018}. LES has been shown to be effective for simulating plumes in previous work in the 
literature, e.g. \citet{pham2007}. The non-dimensional governing equations for velocity $\boldsymbol{u}$ and scalars 
$b, \phi$ including sub-grid-scale (SGS) contributions are
\begin{align}
	\nabla \cdot \widehat{\boldsymbol{u}} &= 0, \label{eq:ns1}\\
	\frac{\mathrm{D} \widehat{\boldsymbol{u}}}{\mathrm{D}t} + \nabla \widehat{p} &=
		\frac{1}{\Rey} \nabla^2 \widehat{\boldsymbol{u}} + \widehat{b} \hat{\boldsymbol{k}} -
		\nabla \cdot \boldsymbol{\tau} + f_w, \label{eq:ns2}\\
	\frac{\mathrm{D}\widehat{b}}{\mathrm{D}t} &= \frac{1}{\Rey\Pr} \nabla^2 \widehat{b} - \nabla \cdot 
        \boldsymbol{\lambda}_b + f_b, \label{eq:ns3}\\
	\frac{\mathrm{D}\widehat{\phi}}{\mathrm{D}t} &= \frac{1}{\Rey\Pr} \nabla^2 \widehat{\phi} - 
        \nabla \cdot \boldsymbol{\lambda}_\phi +
	f_{\phi}, \label{eq:ns4}
\end{align}
where $\widehat{\cdot}$ indicates filtering at the resolved grid scale and $\hat{\boldsymbol{k}}$ is the unit vector 
in the vertical direction. The terms $f_w, \,f_b$ and $f_\phi$ represent the forcing applied to the vertical 
velocity, buoyancy and passive tracer to generate the buoyant plume. The details of this forcing are discussed in 
appendix~\ref{app:A}. The SGS stress tensor $\boldsymbol{\tau}$ has components $\tau_{ij} = \widehat{u_i u_j} - 
\widehat{u}_i \widehat{u}_j$, the SGS buoyancy flux is $\boldsymbol{\lambda}_b =\widehat{\boldsymbol{u}b} - 
\widehat{\boldsymbol{u}} \widehat{b}$ and similarly the SGS tracer flux is $\boldsymbol{\lambda}_{\phi} 
=\widehat{\boldsymbol{u}\phi} - \widehat{\boldsymbol{u}} \widehat{\phi}$. The two dimensionless parameters are the 
Reynolds number and Prandtl number,
\begin{equation}
	\Rey = \frac{F_0^{1/2}}{\nu N^{1/2}}, \quad \Pr = \frac{\nu}{\kappa}, 
\end{equation}
respectively, where $\nu$ is the molecular viscosity and $\kappa$ is the molecular diffusivity for both $b$ and 
$\phi$. The eddy-viscosity model for the deviatoric component of the SGS stress $\boldsymbol{\tau}^d$ and the 
SGS buoyancy and tracer flux are
\begin{equation}
	\tau_{ij}^d = \tau_{ij} - \frac{1}{3}\delta_{ij}\tau_{kk} = -2 \nu_\text{SGS} \widehat{S}_{ij},
	\hspace{2em} \boldsymbol{\lambda}_b = -\kappa^{(b)}_\text{SGS} \nabla \widehat{b},
	\hspace{2em} \boldsymbol{\lambda}_\phi = -\kappa^{(\phi)}_\text{SGS} \nabla \widehat{\phi},
\end{equation}
where $\nu_\text{SGS}, \kappa^{(b)}_\text{SGS}$ and $\kappa^{(\phi)}_\text{SGS}$ are the non-dimensional SGS
viscosity, SGS buoyancy diffusivity and SGS tracer diffusivity respectively, each determined by the AMD
scheme. The term $\widehat{S}_{ij}$ is the non-dimensional shear-rate tensor. The SGS diffusivities and
viscosity may locally exceed the molecular values by several orders of magnitude in regions with intense
turbulence. 

We use DIABLO \citep{taylor2008} to perform three-dimensional LES of the
idealised setup shown in figure~\ref{fig:schematic}. DIABLO evolves the Boussinesq Navier-Stokes equations
\eqref{eq:ns1}--\eqref{eq:ns4} discretised using Fourier modes in the two periodic horizontal directions and
second-order finite differences in the vertical direction. The boundary conditions on the top and bottom boundary are $\partial_z u = \partial_z v = \partial_z b = \partial_z \phi = 0$ and $w = 0$. A third-order Runge-Kutta scheme is used for time-stepping. A $2/3$ dealiasing rule is applied when transforming from Fourier to physical space. We use a cubic domain of side length $L_{\text{domain}}$ with a uniform grid of $512^2 \times 513$ points. The side length is chosen large enough that edge effects are not present and the radially spreading intrusion that forms does not reach the boundary during the simulation. A sponge layer is added in
the top 20\% of the domain (which the simulated plume does not reach), where the velocity is damped towards zero 
and the buoyancy is damped towards the initial background stratification $b(\boldsymbol{x},0) = z$ (non-dimensional), 
to inhibit the reflection of internal gravity waves from the top boundary. Validation of the numerical method discussed here is detailed in appendix~\ref{app:B}.

\begin{table}
	\begin{center}
	\begin{tabular}{cccccccc}
		$\Rey$ & $\Pr$ & $r_0$ & $H$ & $L_{\text{domain}}$ & $L_c$ & $L_p$ & $\tau$ \\[3pt]
		6.29 $\times 10^7$ & 0.70 & 0.20 & 7.97 & 23.9 & 0.80 & 0.40 & 1.00
	\end{tabular}
	\caption{Non-dimensional parameters for the large eddy simulation with $N = 1 \,\mathrm{s}^{-1}$ and $F_0 =
            \bflux \, \mathrm{m}^4\mathrm{s}^{-3}$ discussed from \S\ref{sec:flow} onwards.}
	\label{tab:params}
	\end{center}
\end{table}

Henceforth we refer to a single simulation with parameters which are equivalent to the
experimental setup used by \citet{ansong2010} except for the source buoyancy flux, which is weaker here. The
parameters are given in table~\ref{tab:params} and non-dimensionalised by $N = 1 \,\mathrm{s}^{-1}$ and $F_0 =
\bflux \, \mathrm{m}^4\mathrm{s}^{-3}$. Henceforth, all values stated are non-dimensionalised with respect 
to this choice of $F_0$ and $N$. We also drop the hat notation and refer to the resolved variables unless otherwise noted.

%%%%%%%%%%%%%%%%%%%%%%%%%%%%%%%%%%%%%%%%%%%%%%%%%%%%%%%%%%%%%%%%%%%%%%%%%%%%%%%%%%%%%%%%%%%%%%%%%%%%%%%%%%%%%

\section{Flow \& tracer structure}
\label{sec:flow}

\begin{figure}
	\centering
	\includegraphics[width=\textwidth]{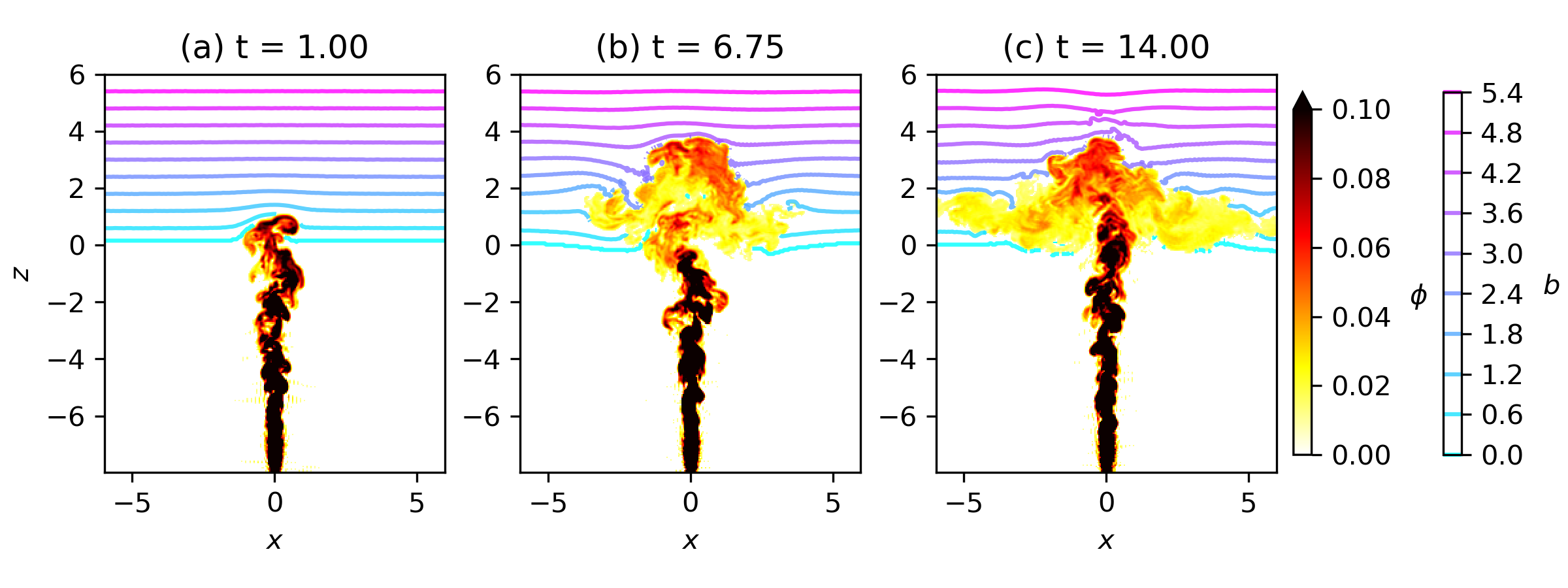}
	\caption{Three stages of the flow evolution, shown as $x-z$ cross-sections of the tracer concentration     
            $\phi$ shown where $\phi$ exceeds $1\%$ of its value on the plume centreline at the source $z=-H$. 
            Buoyancy contours are shown otherwise. Cross-sections are taken at the plume centreline at non-dimensional times $t = 1, 6.75, 14$. From left to right, the panels show the plume during initial penetration, reaching maximum penetration height, and spreading of the intrusion.}
	\label{fig:evol}
\end{figure}

The flow evolution is presented in three vertical cross-sections through the plume centreline in 
figure~\ref{fig:evol}. We identify the plume as regions with tracer concentration $\phi \ge \phi_{\min} \equiv 
10^{-2}$, i.e. we threshold the tracer field at $1\%$ of its value on the plume centreline at the source. In the 
tracer-less environment surrounding the plume we show contours of the buoyancy field. The bottom of 
the stratified layer, above which the buoyancy of the environment becomes non-zero, is indicated by the lowest 
buoyancy contour. 

Figure~\ref{fig:evol}(a) shows initial penetration of the stratified layer by the plume cap. As the plume rises 
through the stratified layer, its upward acceleration decreases as the relative buoyancy between the plume and the 
surrounding environment decreases. Once the environmental buoyancy exceeds that of the plume, the plume decelerates. 
Eventually, the rising fluid reverses direction, or `overturns', and begins to subside from the maximum penetration 
height $\zmax$ (figure~\ref{fig:evol}(b)). As plume fluid subsides, its buoyancy relative to the 
surrounding environment increases until reaching the level of neutral buoyancy $z_n$ 
where the plume fluid forms a radially-spreading intrusion -- see figure~\ref{fig:evol}(c). The
dynamics observed in the simulation agree qualitatively with studies of similar set-ups in the literature, 
for example the experiments detailed in \citet{ansong2010} with an identical setup and similar physical parameters.  

\begin{figure}
	\centering
	\includegraphics[width=\textwidth]{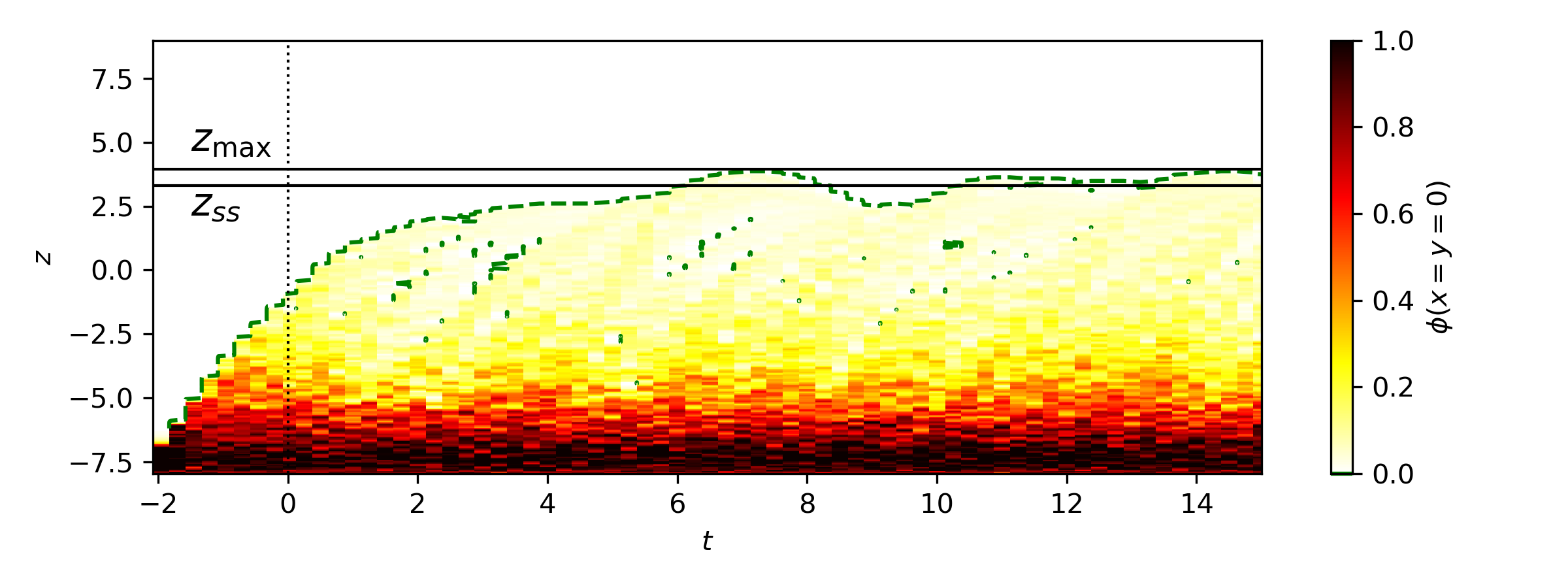}
	\caption{Timeseries of the tracer concentration $\phi(0,0,z,t)$  in the $z-t$ plane at the centreline of the 
            computational domain $x=y=0$. The green dashed contour denotes the plume threshold $\phi = \phi_{\min}$, i.e. where 
            $\phi$ is $1\%$ of its value on the plume centreline. The maximum penetration height $z_{\max}$ and the 
            quasi-steady state height $z_{ss}$ are marked.}
	\label{fig:timeseries}
\end{figure}

The evolution of the maximum height of the plume cap during penetration and the subsequent quasi-steady state is 
visualised as a time-series of tracer concentration on the plume centreline in figure~\ref{fig:timeseries}. As has 
been noted in the literature, the maximum height of the plume tends to oscillate around a quasi-steady state height 
$z_{ss}$ \citep{turner1966} but, to our knowledge, the mechanism setting the frequency of this oscillation (often 
referred to as `plume bobbing') is not well understood (e.g. \citet{ansong2010}). In the simulation considered here, 
the quasi-steady state height $z_{ss}$ is close to the maximum penetration height $z_{\max}$ and the oscillation is 
weak. For convenience, we will use $z_{\max}$ to refer to the maximum height of the plume. The maximum 
penetration height $\zmax$ determines the maximum height at which plume fluid can mix with the 
environment \citep{ansong2008}, meaning the initial buoyancy at the maximum penetration height, $b = z_{\max}$, 
represents a plausible constraint on the maximum buoyancy accessible for mixing with the plume. 
However, this constraint can occasionally be exceeded when plume fluid subsiding from the plume cap pulls 
very buoyant environmental fluid downwards (see figure~\ref{fig:evol}(b) to the left of the plume cap). 
Here we find $\zmax = \zmaxval$ which agrees with experimental estimates of the maximum penetration height in the 
literature, e.g. $\zmax \approx 3.8$ for a plume with a round source \citep{list1982}.

Internal gravity waves across a range of frequencies are generated during the penetration process. These waves
are visible as small amplitude, long wavelength undulations in the buoyancy contours above $z \approx 4$ in 
figure~\ref{fig:evol}(c). Any influence of internal gravity waves on mixing in this flow will be present in the analyses, 
but it is beyond the scope of this paper to determine the particular contribution of these waves to mixing.

In the uniform layer, the buoyancy and tracer evolve identically up to a linear factor, i.e. the undiluted plume fluid 
entering the stratified layer has a linear relationship between $b$ and $\phi$ at each point. This follows from the 
self-similarity of the buoyancy and tracer concentration profiles in the steady state plume that penetrates the 
stratified layer (see appendix~\ref{app:B}, figure~\ref{fig:profiles}). The radial profiles for $b$ and $\phi$ are 
both Gaussian with the same width but different amplitudes, hence $b \propto \phi$. After penetrating the stratified 
layer, plume fluid with non-zero buoyancy and tracer concentration mixes with tracer-less environmental fluid and hence 
the buoyancy and tracer evolve differently. This effect can be quantified using a 
tracer probability density function (PDF) in buoyancy coordinates $\tilde{\phi}(b; t)$. The PDF is calculated within 
the stratified layer only. The value of the PDF $\tilde{\phi}(b; t)$ is calculated as the total tracer 
with buoyancy within a range $b$ to $b + \mathrm{d}b$ in the stratified layer, normalised by the total tracer in the 
stratified layer $\phi_T(t) = \sum_V \phi(\boldsymbol{x},t)\Delta V$, where $V$ is the stratified layer and $\Delta V$ 
is the grid-cell volume. The definition of $\tilde{\phi}$ is such that $\sum_B \tilde{\phi}(B; t) = 1$.

\begin{figure}
	\centering
	\includegraphics[width=\textwidth]{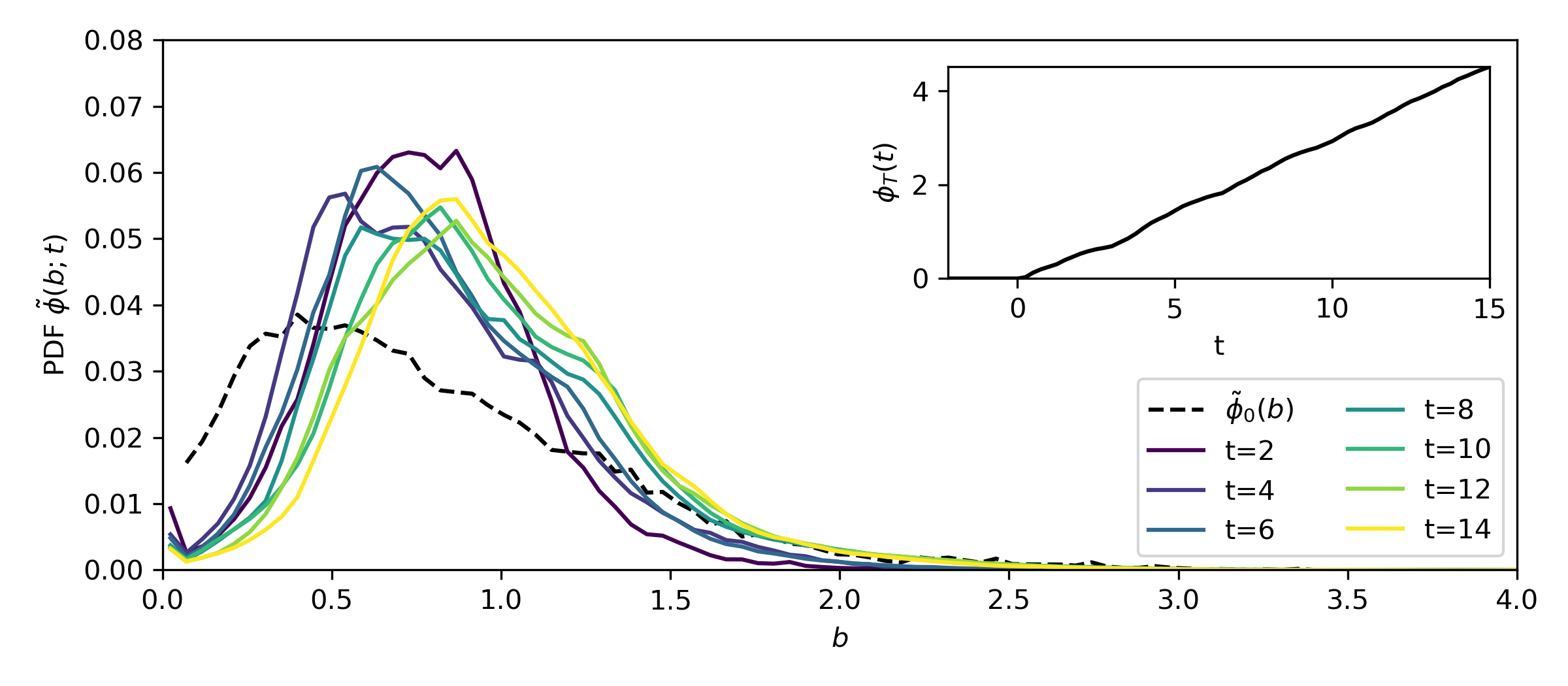}
	\caption{Probability density function $\tilde{\phi}(b, t)$ of tracer as a function of buoyancy 
        $b$ in the stratified layer at fixed time intervals post-penetration shown as coloured lines. The black dashed 
        line shows the time-averaged pre-penetration PDF $\tilde{\phi}_0(b)$, calculated with $V$ chosen as the 
        pre-penetration region indicated in figure~\ref{fig:schematic} and time-averaged. The pre-penetration PDF 
        $\tilde{\phi}_0(b)$ shows the tracer-buoyancy relationship within the plume prior to penetrating the 
        stratified layer. Differences between $\tilde{\phi}_0$ and $\tilde{\phi}(b,t)$ represent the effect of mixing. 
        Total tracer in the stratified layer $\phi_T(t)$ shown inset.}
	\label{fig:tb_dist}
\end{figure}

Figure~\ref{fig:tb_dist} shows $\tilde{\phi}(b;t)$ in the stratified
layer at fixed time intervals post-penetration. The total tracer in the stratified layer $\phi_T(t)$ 
is shown inset. The approximately linear increase in $\phi_T$ with time suggests a relatively uniform input of 
tracer to the stratified layer, carried by the penetrating plume. Owing to the self-similar nature of the 
penetrating plume, we expect the tracer that enters the stratified layer to have a fixed PDF (with some small 
variation). This \emph{pre-penetration} PDF $\tilde{\phi}_0$ can be estimated using a domain $V$ chosen as the 
pre-penetration region shown in figure~\ref{fig:schematic}. This region is a thin layer with (non-dimensional) depth 
$1/2$ below the bottom of the stratified layer. The pre-penetration PDF, shown as a black dashed line in 
figure~\ref{fig:tb_dist}, represents the tracer PDF in the plume just before it penetrates the stratified layer. 
Without mixing, $\tilde{\phi}$ in the stratified layer would match the pre-penetration PDF. Mixing during the penetration 
process manifests as changes in the tracer PDF when compared with the pre-penetration PDF.

Evolution of the post-penetration PDF and changes compared with the pre-penetration tracer PDF highlight two mixing processes 
during penetration: mixing within the plume during penetration, and mixing between the plume and environment. Where plume 
fluid carries a large tracer concentration and mixes with the more buoyant surroundings, the positive tail of the tracer PDF 
increases. This is particularly evident after $t=\tmaxval$ when the plume has reached 
$\zmax=\zmaxval$, at which point very large values of buoyancy in the environment become accessible \emph{and} large 
tracer concentrations at the centre of the plume are exposed to the environment as plume fluid overturns. At the edges 
of the plume where tracer concentration is smallest, mixing with the environment again moves tracer from lower to 
higher values of buoyancy and therefore the PDF decreases where $b$ is small. This effect is supplemented by mixing 
within the plume, which acts to homogenise the large tracer concentration and buoyancy at the centre of the plume with 
the lower tracer concentration and buoyancy at the edge of the plume. This acts to narrow the PDF and hence decrease the PDF at large and small values of buoyancy but the effect is only evident before the plume reaches $\zmax$ at $t=\tmaxval$ and 
accesses much larger values of buoyancy. At late times, most tracer lies in the spreading intrusion at the neutral 
buoyancy height $z_n$, which coincides with the peak in the tracer PDF.

The buoyancy range of the tracer PDF is determined by the maximum penetration height of the plume as well as the 
rapidity of the mixing between the plume and environment occurring in the plume cap. If fluid quickly 
reaches $\zmax$ and subsides before substantial mixing with the environment occurs, only small amounts of the more 
buoyant environment are entrained and therefore the increase in the PDF at large values of buoyancy is modest compared 
with a scenario where plume fluid stalls during overturning and significant mixing with the environment occurs. In 
figure~\ref{fig:tb_dist}, the tracer PDF extent is $b \approx 3$ whilst the environmental buoyancy at $\zmax$ is 
approximately $\left. b \right|_{\zmax} \approx \zmaxval$. This suggests the mixing timescale is slow compared to the 
dynamical timescale, i.e. mixing between the largest tracer concentrations first exposed during overturning and the 
environment is slow and continues during subsidence, where the buoyancy of the environment decreases. 

The tracer PDF hints at competing effects of mixing within the plume and between the plume and the 
environment. Crucially, the buoyancy and tracer fields are mixed in different ways owing to the linearly increasing 
buoyancy and vanishing tracer concentration in the linearly stratified environment. Whilst changes in the tracer PDF 
considered here demonstrate the overall effect on the relation between tracer and buoyancy, it is difficult to extract 
information on the intensity of mixing between plume and environmental fluid and the specific buoyancy and tracer 
characteristics of the fluid parcels that mix. Furthermore, the tracer PDF $\tilde{\phi}(b;t)$ does not give information on 
the volume of fluid parcels at a given buoyancy; a peak in the tracer PDF may represent 
relatively few fluid parcels carrying large tracer concentrations or many fluid parcels carrying small 
amounts of tracer. The distinction is important, since the former can result in stronger gradients upon which 
diffusion acts and therefore more effective diapycnal transport of tracer.

\section{Buoyancy-tracer volume distribution}
\label{sec:VD}

The probability distributions of tracer concentration discussed in section~\ref{sec:flow} isolate the 
\emph{irreversible} transport that results from turbulent mixing. The turbulent mixing of fluid parcels can be 
considered a two-step process (e.g. \citet{wykes2014}), composed of stirring and molecular diffusion. Whilst 
stirring strengthens tracer gradients across buoyancy surfaces, it is -- in principle -- a reversible process. However, 
molecular diffusion results in irreversible changes to the buoyancy and tracer characteristics of fluid parcels and hence 
changes the tracer distribution.

Here, we use the distribution of volume in buoyancy-tracer space to diagnose mixing in the stratified layer. That is, we map from 3D physical space to a 2D phase space by using the buoyancy and tracer concentration fields to quantify the volume of plume fluid in the stratified layer with each value of $b$ and $\phi$. The total physical volume of plume fluid represented in the distribution changes in time and we do not normalise the distribution to form a PDF. Omitting this normalisation simplifies the interpretation of the distribution and its governing equation. The buoyancy-tracer volume distribution formalism presented here builds on previous density-tracer joint PDF formulations presented by \citet{plumb2007} and \citet{penney2020}. 

\subsection{Definition \& properties}
\label{sec:defn}
We define the \emph{volume distribution} $W(B, \Phi; t)$ in buoyancy-tracer space such that the volume of fluid in a 
fixed volume $V$ with $B < b(\boldsymbol{x}, t) < B + \mathrm{d}B$ and $\Phi < \phi(\boldsymbol{x}, t) < \Phi + 
\mathrm{d}\Phi$ is given by $W(B, \Phi; t)\mathrm{d}B\mathrm{d}\Phi$. This may be defined as
\begin{equation}
	W(B, \Phi; t) = \int_V \delta(b(\boldsymbol{x}, t) - B) \delta(\phi(\boldsymbol{x}, t) - \Phi) \,
	\mathrm{d}V,
	\label{def:W}
\end{equation}
where $\delta(\cdot)$ is the Dirac delta function with the inverse dimension of its argument. Henceforth we 
choose the volume $V$ to be the stratified layer. An evolution equation for $W$ can be obtained using the governing 
equations for $b$ and $\phi$. See appendix~\ref{app:C} for a full derivation. We have
\begin{equation}
	\frac{\partial W}{\partial t} = - \bnabla_{(B, \Phi)} \cdot \boldsymbol{F} + S,
	\label{eq:VDbudget}
\end{equation}
where $\boldsymbol{F}(B, \Phi; t)$ is \emph{mixing flux distribution} and $S(B, \Phi; t)$ is the \emph{source distribution}. The mixing flux distribution $\boldsymbol{F}$ is a vector in buoyancy-tracer space with components formed from the 
volume-weighted average of the non-advective terms $\dot{b}$ and $\dot{\phi}$ in \eqref{eq:ns3} and \eqref{eq:ns4} respectively, representing the flux of $W$ in buoyancy-tracer space due to mixing and is defined as
\begin{equation}
	\boldsymbol{F}(B, \Phi; t) = (F_b, F_\phi) = \int_V (\dot{b}, \dot{\phi})\,\delta(b(\boldsymbol{x}, t) - B)
	\delta(\phi(\boldsymbol{x}, t) - \Phi) \, \mathrm{d}V, \label{def:F}
\end{equation}
where $\dot{b} = (\Rey \Pr)^{-1} \nabla^2 b - \nabla \cdot \boldsymbol{\lambda}_b$ and $\dot{\phi} =    
(\Rey\Pr)^{-1}\nabla^2 \phi - \nabla \cdot \boldsymbol{\lambda}_\phi$. Note that the plume forcing terms $f_b$ and $f_\phi$ are excluded from $\dot{b}$ and 
$\dot{\phi}$ since the forcing vanishes in the stratified layer. The source distribution $S$ represents a source 
or sink of $W$ due to boundary fluxes across $\partial V$,
\begin{equation}
	S(B, \Phi; t) = \int_{\partial V} \boldsymbol{u}\cdot\boldsymbol{n}\, \delta(b(\boldsymbol{x}, t) - B)
	\delta(\phi(\boldsymbol{x}, t) - \Phi) \, \mathrm{d}A, \label{def:S}
\end{equation}
where $\boldsymbol{u}$ is the velocity in physical space and $\boldsymbol{n}$ is the \emph{inward}
normal on the boundary $\partial V$ of $V$. Since we are considering a flow upwards into $V$, 
$\boldsymbol{u}\cdot\boldsymbol{n}$ is positive and $S$ acts as a source of $W$. Note that whilst $S$ 
represents the effect of fluxes across the boundary $\partial V$ in physical space, it is distributed 
in buoyancy-tracer space. Note that \eqref{eq:VDbudget} contains no terms in which advection plays an 
explicit role except for the source term -- which captures advection through the domain 
boundary -- representing the fact that $W$ remains unchanged under advection within the domain.

Turbulent mixing redistributes volume in buoyancy-tracer space, which results in 
changes to $W$ via the mixing flux term $-\nabla_{(B,\Phi)} \cdot \boldsymbol{F}$. The change in
$W$ at a point $(B, \Phi)$ in buoyancy-tracer space as a result of turbulent mixing up to time $t$ is 
therefore
\begin{equation}
	M(B, \Phi; t) = -\int_0^t \nabla_{(B, \Phi)} \cdot \boldsymbol{F}(B, \Phi; t')\,
	\mathrm{d}t' = W(B, \Phi; t) - \int_0^t S(B, \Phi; t') \, \mathrm{d}t', \label{def:M}
\end{equation}
such that $M(B, \Phi; t)\mathrm{d}B \mathrm{d}\Phi$ is the change in volume of fluid with 
$B < b(\boldsymbol{x}, t) < B + \mathrm{d}B$ and $\Phi < \phi(\boldsymbol{x}, t) < \Phi +
\mathrm{d}\Phi$ up to time $t$ due to mixing. Therefore $M$ represents the
integrated effect of the mixing flux $\boldsymbol{F}$ and we refer to $M$ as the \emph{net mixing effect 
distribution}. The second equality in \eqref{def:M} follows from time-integrating 
\eqref{eq:VDbudget} and noting that $W(B, \Phi; t=0) = 0$ since there is no tracer in the initial stratified layer. 
Hence, $M$ can be interpreted as a cumulative measure of the changes to $W$ relative to the time-integrated source 
distribution, i.e. the changes in the volume distribution that arise solely from mixing. The final term in \eqref{def:M}, 
which we refer to as the \textit{cumulative source distribution}, represents the volume of fluid with buoyancy $B < b < B + 
\mathrm{d}B$ and tracer concentration $\Phi < \phi < \Phi + \mathrm{d}\Phi$ that has entered the stratified layer up to time 
$t$. The volume distribution $W \geq 0$ and the cumulative source distribution is also positive assuming there is a flow into 
$V$ only. However, $M$ can be positive or negative depending on the relative sizes of the volume distribution and the 
cumulative source distribution. 

The net mixing effect distribution 
$M(B, \Phi; t)$ is \emph{positive} in buoyancy-tracer space where more volume is present at time $t$ than has entered 
the stratified layer up to time $t$, i.e. there is a net gain in the volume of fluid with buoyancy $B$ and tracer 
concentration $\Phi$ due to mixing. Correspondingly, $M(B, \Phi; t)$ is \emph{negative} where more volume has entered 
the stratified layer up to time $t$ with buoyancy $B$ and tracer concentration $\Phi$ than currently exists at time 
$t$, i.e. there is a net loss in the volume of fluid with buoyancy $B$ and tracer concentration $\Phi$ due to mixing. 
The value of $M$ therefore indicates the transfer of volume within $W$ due to mixing; fluid leaves regions of 
buoyancy-tracer space with $M < 0$ and enters regions with $M > 0$. 

To summarise, the distributions $W, S, \boldsymbol{F}$ and $M$ together describe the flow in terms of its 
effect on buoyancy-tracer space. The volume distribution $W$ is an instantaneous representation of the amount of fluid 
within the stratified layer with given ranges of values of buoyancy and tracer concentration. Large values of $W$ 
indicate large volumes of fluid with a narrow range of $b$ and $\phi$, though the fluid parcels corresponding to this 
range are not necessarily co-located in physical space. The source distribution $S$ represents the volume distribution 
of fluid that enters the stratified layer from the uniform layer. In the absence of mixing, $W$ would be equivalent to 
the time integral of $S$. The mixing flux distribution $\boldsymbol{F}$ represents the redistribution of 
fluid in buoyancy-tracer space due to mixing. The net mixing flux distribution $M$ captures the change in 
$W$ relative to time-integrated $S$ via $\boldsymbol{F}$ and indicates where there is accumulation or loss of volume 
due to mixing.

\begin{figure}
	\centering
	\includegraphics[width=\textwidth]{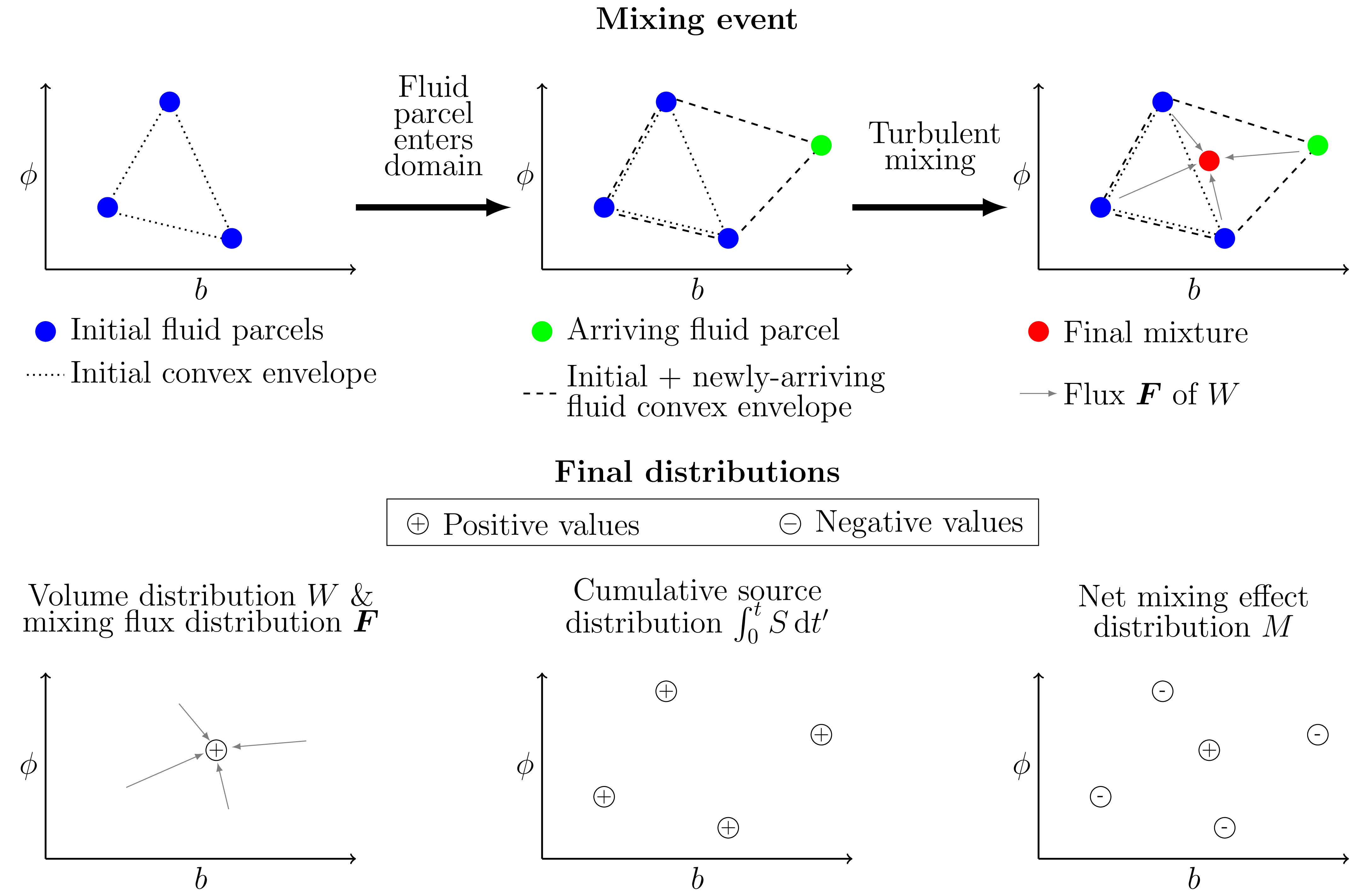}
	\caption{Schematic diagram of the effect in buoyancy-tracer space of an idealised turbulent mixing event 
 between a set of discrete fluid parcels that have entered the stratified layer and a fluid parcel that later enters the 
 stratified layer. The top row illustrates the convex envelope property of the volume distribution $W$, which implies 
 that a mixture of a set of fluid parcels lies within the smallest envelope that contains the distribution of the 
 fluid parcels that are mixed together. The distributions $W, \int S\, \mathrm{d}t$ and $M$ following the idealised 
 mixing event are shown on the bottom row. Positive (arbitrary) values of each distribution are indicated by 
 a circled $+$, negative (arbitrary) values are indicated by a circled $-$. The mixing flux 
 distribution vectors $\boldsymbol{F}$ are indicated by grey arrows.}
	\label{fig:vd_properties}
\end{figure}

\subsection{Idealised example}
The effect of an idealised turbulent mixing event on buoyancy-tracer space is illustrated in 
figure~\ref{fig:vd_properties}. The top row shows the volume distribution containing three initial fluid parcels (blue 
points) which have entered the stratified layer. As turbulent stirring brings these fluid parcels 
together, another fluid parcel (green point) enters the domain and all four fluid parcels mix. The resulting mixed 
fluid parcel (red point) is a volume-weighted average of the fluid parcels involved in the mixing event. The bottom 
row shows the distributions after the mixing event. The volume distribution $W$ is non-zero (and positive) only where 
the final mixture lies in buoyancy-tracer space and the direction of the mixing flux vectors $\boldsymbol{F}$ 
indicates the redistribution of volume. The cumulative source distribution $\int_0^t S \, \mathrm{d}t'$ is positive at the 
values of $b$ and $\phi$ where the fluid parcels entered the domain and vanishes elsewhere. The net mixing effect 
distribution $M$ is negative at these points, as volume has been lost, and positive where the mixed fluid parcel lies as 
volume has been gained. These principles can be used to understand the mixing processes in the physical flow that 
result in changes in the distribution in buoyancy-tracer space. Whilst it is not possible 
to isolate distinct fluid parcels that are mixing at any one time, we can identify physical regions of the flow 
that are subject to turbulent mixing and isolate the corresponding regions of buoyancy-tracer space.

Turbulent mixing acts to homogenise the buoyancy and tracer concentration of fluid parcels. Provided 
the molecular diffusivities of buoyancy and tracer are equal, a mixture of two fluid parcels lies on a line between 
the two parcels in buoyancy-tracer space \citep{penney2020}. Therefore, as the buoyancy-tracer volume distribution $W$ 
evolves, it is constrained to lie within its own past convex envelope, i.e.\ the smallest convex set that contains all 
non-zero points of the distribution. As illustrated in figure~\ref{fig:vd_properties}, this convex envelope must include 
fluid that enters the domain during the mixing process. The convex envelope of the initial volume distribution is 
indicated by the dotted envelope and the dashed envelope indicates the convex envelope including newly-arriving fluid 
parcels. We emphasise that the final mixed fluid parcel is contained within the convex envelope of initial and newly-arriving 
fluid, but not necessarily within the convex envelope of the initial fluid only. The principle of homogenising 
fluid parcels illustrated in figure~\ref{fig:vd_properties} can be generalised to continuous mixing of fluid in a flow, in which case the convex envelope constraint applies to the volume distribution as a whole. This implies that 
in the absence of sources the convex envelope reduces over time and converges towards some compact distribution 
\citep{penney2020}. In the setup we consider, fluid entering the stratified layer causes the extremes of the distribution 
to persist, whilst turbulent mixing acts to continuously shift the buoyancy-tracer characteristics of fluid towards an 
accumulation region in buoyancy-tracer space.

\subsection{Numerical implementation}
\label{sec:numimp}
To examine the numerical simulation detailed in \S2 and \S3, we use a discrete formulation of the
buoyancy-tracer volume distribution introduced in \S\ref{sec:defn}. We choose the domain $V$ to be tracer-containing 
plume fluid within the stratified layer. The stratified layer initially corresponds to the volume $z \ge 0$. However, 
the plume can perturb the bottom of the stratified layer slightly below $z=0$. We therefore define the domain as the 
region where $-1 \le z \le L_z$,  $\phi > \phi_{\min}$, and $b > 0$. As a consequence, the `reservoir' of 
environmental fluid where $\phi = 0$ is excluded. In interpreting results, we therefore consider the boundary $\phi = 
\phi_{\min}$ as a source where volume can enter the distribution from the environment. The entrainment of 
environmental fluid across this boundary in buoyancy-tracer space into the volume distribution is discussed in 
\S\ref{sec:entrainment} and illustrated in figure~\ref{fig:VDcorr}.

The buoyancy and tracer domains are subdivided into $N_b$ and $N_{\phi}$ equally sized bins of size $\delta B = 
(b_{\max} - b_{\min})/N_b$ and $\delta \Phi = (\phi_{\max} - \phi_{\min})/N_\phi$ respectively.
We choose $b_{\min} = 0$ and $b_{\max} = 4$ since the largest accessible buoyancy is related to the 
maximum penetration height \citep{ansong2008} which is experimentally predicted to be $\zmax \le 4$ \citep{list1982}.
To accommodate fluctuations in tracer concentration, we choose $\phi_{\max}$ to be larger than the tracer 
concentration on the plume centreline $2\phi_m(0)$ at penetration height $z=0$ using the profile predicted by the 
\citet{mtt} plume equations, $\phi_m(z)$, defined in \eqref{eq:plume3} (see appendix~\ref{app:A} for details). We use 
$\phi_{\min} = 10^{-2}$, consistent with the plume threshold introduced in \S\ref{sec:flow}. Henceforth we use $N_b = 
N_\phi = 256$.

Denoting the centre of a given bin as $(B_i, \Phi_j)$, the associated value of the volume distribution
is computed as
\begin{equation}
	W_{ij}(t) = \sum_V I_{ij}(\boldsymbol{x},t) \Delta x \Delta y \Delta z
\end{equation}
where the sum is over all grid points within the domain $V$, $\Delta x, \Delta y, \Delta z$ are the grid-cell widths,
and the indicator $I_{ij}(\boldsymbol{x},t)$ is defined as
\begin{equation}
	I_{ij}(\boldsymbol{x}, t) = \begin{cases}
		1 & \left(b(\boldsymbol{x},t) - B_i, \phi(\boldsymbol{x},t) - \Phi_j\right) \in 
			\bigl( -\frac{1}{2}\delta B, \frac{1}{2}\delta B \bigr] \times 
			\bigl( -\frac{1}{2}\delta \Phi, \frac{1}{2}\delta \Phi \bigr], \\
			0 & \text{otherwise}.
		\end{cases}
		\label{eq:indicator}
\end{equation}
The value of $W_{ij}(t)$ therefore represents the total volume within $V$ where the buoyancy lies within
$\delta B/2$ of $B_i$ and the tracer concentration lies within $\delta \Phi/2$ of $\Phi_j$. Note that
in the continuous formulation, the volume distribution $W(B, \Phi; t)$ defined by \eqref{def:W} must be
integrated over $B$ and $\Phi$ to yield a volume, whilst $W_{ij}(t)$ itself has dimensions of volume and need
only be summed over $i$ and $j$. The continuous and discrete formulations coincide in the 
limit $\delta B, \delta \Phi \to 0$, such that
\begin{equation}
	\lim_{\delta B, \delta \Phi \to 0} \frac{W_{ij}(t)}{\delta B \delta \Phi} = W(B_i, \Phi_j; t). \label{eq:limit}
\end{equation}
The equivalence \eqref{eq:limit} also applies to the discrete mixing flux distribution $\boldsymbol{F}_{ij}(t)$, the discrete source distribution $S_{ij}(t)$ and the discrete net mixing effect distribution $M_{ij}(t)$ defined by
\begin{align}
    \boldsymbol{F}_{ij}(t) &= (F^b_{ij}(t), F^\phi_{ij}(t)) = \sum_{V} I_{ij}(\boldsymbol{x}, t) (\dot{b},
	\dot{\phi}) \Delta x \Delta y \Delta z, \label{eq:discreteF}\\
    S_{ij}(t) &= \sum_{\partial V} I_{ij}(\left.\boldsymbol{x}\right|_{z=-1}, t)
	w(\left.\boldsymbol{x}\right|_{z=-1}, t) \Delta x \Delta y, \label{eq:discreteS}\\
    M_{ij}(t) &= W_{ij}(t) - \sum_{t'} S_{ij}(t') \Delta t', \label{eq:discreteM}
\end{align}
where $\Delta t'$ is the simulation time step. In \eqref{eq:discreteF}, $\dot{b}$ and $\dot{\phi}$ are the non-advective 
terms in the scalar evolution equations \eqref{eq:ns3}, \eqref{eq:ns4} of $b, \phi$ respectively, as 
defined in \S\ref{sec:defn}. In \eqref{eq:discreteS} we have used the fact that $\boldsymbol{n} = 
\hat{\boldsymbol{k}}$ on the bottom boundary of the domain $V$.

\subsection{Results}
\label{sec:results}

The discrete formulation of the distributions given in \S\ref{sec:numimp} provides an approximation to the continuous 
formulation and is presented in all figures shown below. However, the interpretation is the same as the 
continuous formulation and we will refer to the continuous formulation in all discussions. Quantities derived from the 
distributions are given in both continuous and discrete forms for completeness. In defining the discrete and 
continuous formulations we use the arguments $B$ and $\Phi$, which represent values of buoyancy and tracer 
concentration respectively. We treat $W, \, \boldsymbol{F}, \, S$ and $M$ as functions of $b$ and $\phi$ 
to aid clarity, e.g. $W(b, \phi; t)$, with the interpretation that $b$ and $\phi$ represent values of buoyancy and 
tracer concentration found in the flow in the same way as $B$ and $\Phi$ in \S\ref{sec:defn} and \S\ref{sec:numimp}.

\begin{figure}
	\centering
	\includegraphics[width=\textwidth]{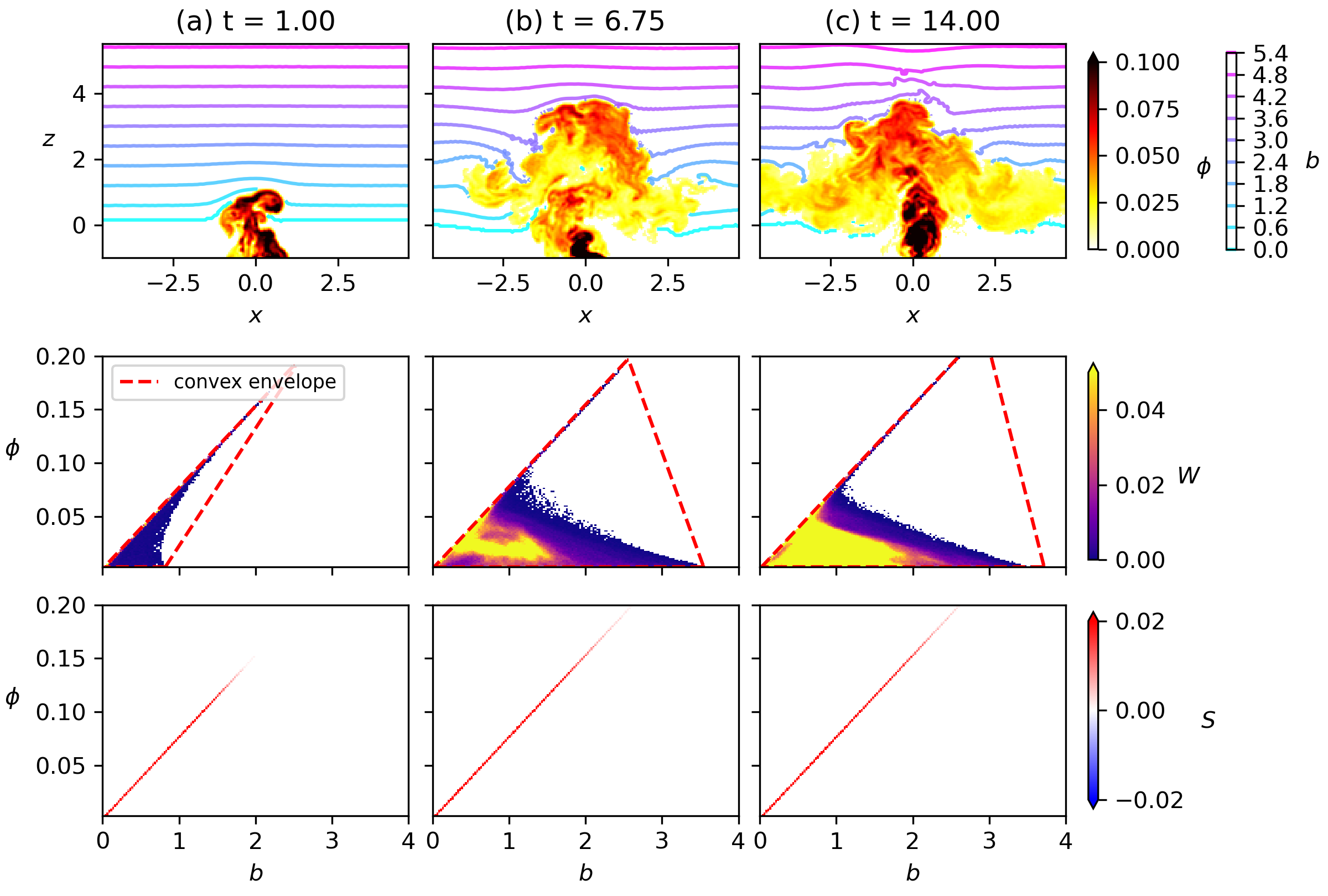}
	\caption{Three instantaneous snapshots showing the evolution of the buoyancy-tracer volume distribution 
 $W(b, \phi; t)$ (middle) and source distribution $S(b, \phi; t)$ (bottom) at non-dimensional times $t=1, 6.75, 14$ 
 corresponding with figure~\ref{fig:evol}. The convex envelope of the volume distribution $W$ at time $t$ is shown as 
 a red dashed line in the middle panel. To aid interpretation, we also show $x-z$ cross-sections of the tracer
 concentration and buoyancy contours, as in figure~\ref{sec:flow} (top)}
	\label{fig:VD}
\end{figure}

Figure~\ref{fig:VD} shows the buoyancy-tracer volume distribution $W(b, \phi; t)$ (middle row), the source 
distribution $S(b, \phi; t)$ (bottom row), and $x-z$ cross-sections of the tracer concentration field and buoyancy 
contours (top row). These results are shown at three snapshots corresponding with stages of the flow evolution as in 
figure~\ref{fig:evol}. The distributions are shown only where non-zero, i.e. regions of buoyancy-tracer space which 
are not coloured indicate that there is no fluid with buoyancy and tracer
concentration in that range. In each snapshot of $W$, the red dashed lines show the convex envelope that constrains 
the evolution of the volume distribution. As seen in the figure, the source distribution lies within the convex 
envelope of $W(b, \phi; t)$. Furthermore, as the plume rises and accesses more buoyant fluid in the surrounding 
environment, the convex envelope is extended along the $\phi=0$ axis as new environmental fluid becomes accessible via 
mixing.

\begin{figure}
	\centering
	\includegraphics[width=.6\textwidth]{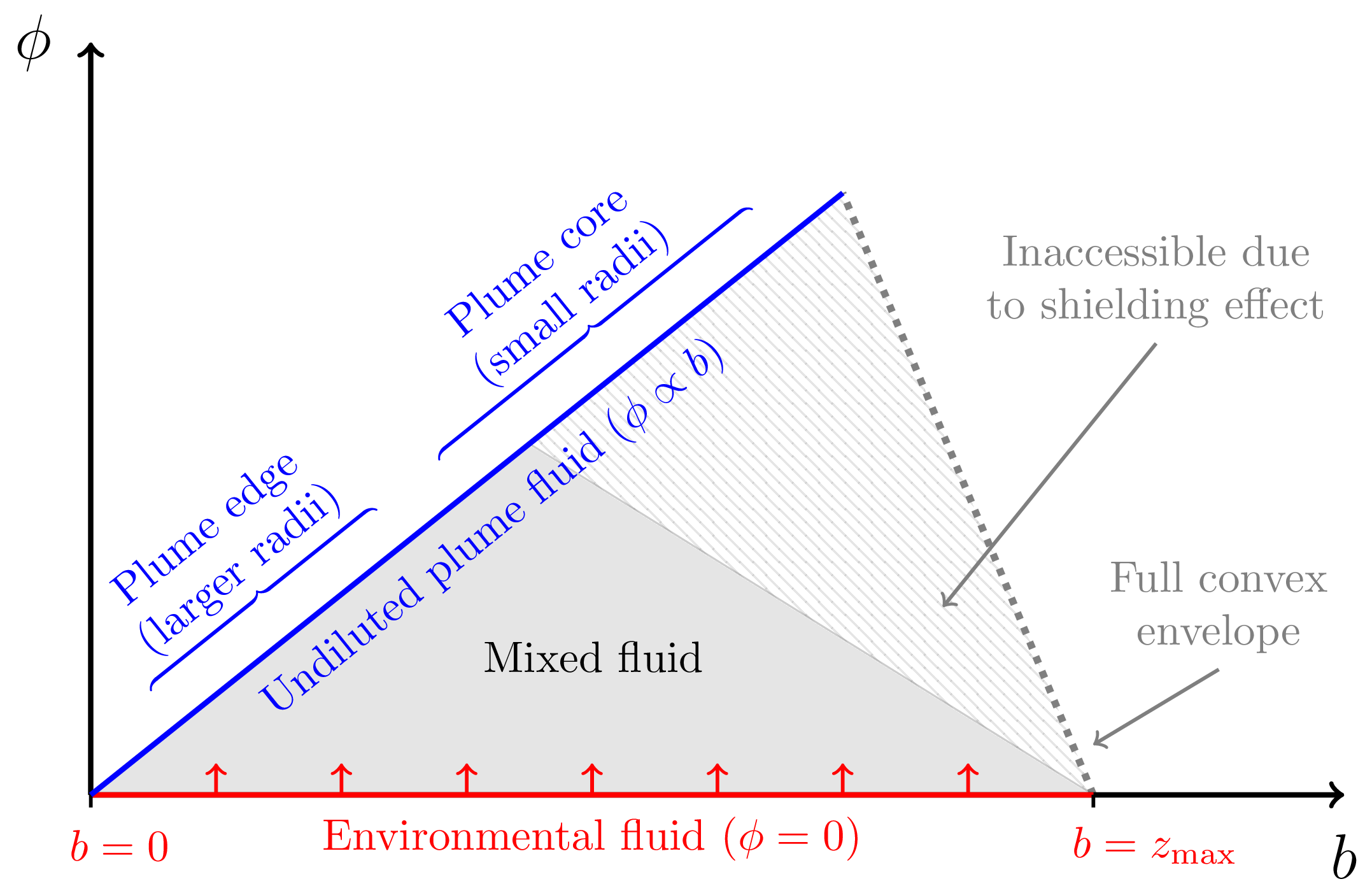}
	\caption{Schematic diagram illustrating the correspondence between regions in physical space and regions
in buoyancy-tracer space. Environmental fluid is represented by the $b$ axis where $\phi=0$ (red), between the bottom 
of the stratified layer where $b=0$ and the maximum penetration height where $b=\zmax$. Undiluted plume 
fluid lies on the source line where $\phi \propto b$ (blue), with large $b$ and $\phi$ in the core of the plume and 
small $b$ and $\phi$ towards the edges. Mixed fluid lies between these two lines, within their 
convex envelope (dotted gray). Some regions of the convex envelope (hatched area) are inaccessible due to the shielding effect of the 
plume edge (where $\phi$ is non-zero but small) and the intrusion surrounding the rising plume. Volume entering the distribution 
due to entrainment of environmental fluid is indicated by the red arrows.}
	\label{fig:VDcorr}
\end{figure}

The results shown in figure~\ref{fig:VD} illustrate how the volume distribution captures the dynamics and mixing 
processes at each stage of the flow evolution. We first note that the source distribution $S(b, \phi; t)$ takes 
positive values only, since there is only a flow \emph{into} the stratified layer. Furthermore, $S$ is non-zero only on a line through the origin as expected from the linear relationship between $b$ and $\phi$ in the rising plume. 
We refer to this as the \emph{source line}. Given that the convex envelope 
of a set of points on a line segment is the same line segment, mixing of undiluted plume fluid within the plume only 
redistributes fluid on the source line. When undiluted plume fluid mixes with the surrounding environment, it is moved away from the source line. This offers a clear distinction between undiluted and mixed plume 
fluid, as illustrated schematically in figure~\ref{fig:VDcorr}. In the buoyancy-tracer volume distribution $W$ shown 
on the middle row of figure~\ref{fig:VD}, fluid appearing away from the source line 
therefore represents a mixture of plume and environmental fluid. Further information on the regions of the undiluted 
plume that mix with the environment is gained by noting that, owing to the Gaussian profiles of the plume 
pre-penetration, $b$ and $\phi$ are larger near the centreline of the plume and smaller towards the edge of the plume 
(see figure~\ref{fig:VD}(c)). Hence fluid near the `plume edge' lies nearest the origin on the source line whilst 
fluid in the `plume core' lies at the extreme end of the source line.
 
Figure~\ref{fig:VD}(a) shows the plume shortly after penetrating the stratified layer and before reaching its
maximum penetration height. At this stage, only the edges of the plume are exposed to the environment in the
lower part of the stratified layer which has a relatively small buoyancy. We therefore find volume appearing in a small
region of the convex envelope closest to the origin in buoyancy-tracer space. In figure~\ref{fig:VD}(b), the
plume has reached its maximum penetration height and undiluted plume fluid in the plume core, which has until
this stage been shielded from the environment by the edges of the plume, overturns and becomes exposed to
environmental fluid near $\zmax$ with relatively large buoyancy. The volume distribution at this time covers a wider 
range of $b$ and $\phi$ as environmental fluid with a large range of buoyancy is able to mix with much of the 
undiluted plume fluid lying on the source line. However, note that we do not see mixing between the extreme end of 
the source line and the environment (see hatched region in figure~\ref{fig:VDcorr}). As undiluted plume fluid in 
the core of the plume rises, turbulent mixing acts to homogenise the buoyancy and tracer concentration within the 
source distribution, reducing the largest values of $b$ and $\phi$ near the centreline of the plume. Hence when this 
fluid is first exposed to the environment in the plume cap, it lies closer to the middle of the source line. We refer 
to this as the shielding effect. Note that the large values of $b$ and $\phi$ on the source line appear to persist, and may increase, due to new undiluted plume fluid entering the stratified layer.

Figure~\ref{fig:VD}(c) shows the quasi-steady state plume where there is a continuous input of undiluted plume 
fluid along the source line $\phi \propto b$, mixing between the plume and environment up to values of 
buoyancy corresponding to heights near $\zmax$ and an accumulation of fluid at lower values of buoyancy and tracer 
concentration corresponding to the radially spreading intrusion. The intrusion dominates the total volume of the 
plume at late times and is represented by the region of large $W$ at intermediate values of buoyancy and tracer 
concentration. Note that the intrusion enhances the shielding effect by preventing undiluted plume fluid from 
accessing the environment before reaching the plume cap. 

\subsection{Quasi-steady state}
\label{sec:qss}

In this flow, quasi-steady state refers to the long-term behaviour established once an intrusion has formed. In this state, undiluted plume fluid continuously arrives in the stratified layer, mixes with the environment, and
accumulates in the intrusion. This means that in quasi-steady state the volume of undiluted plume fluid in the
stratified layer remains roughly constant whilst the full plume volume (i.e. the sum of the rising undiluted plume, 
plume cap and intrusion), and in particular the volume of the intrusion, increases monotonically.

\begin{figure}
	\centering
	\includegraphics[width=\textwidth]{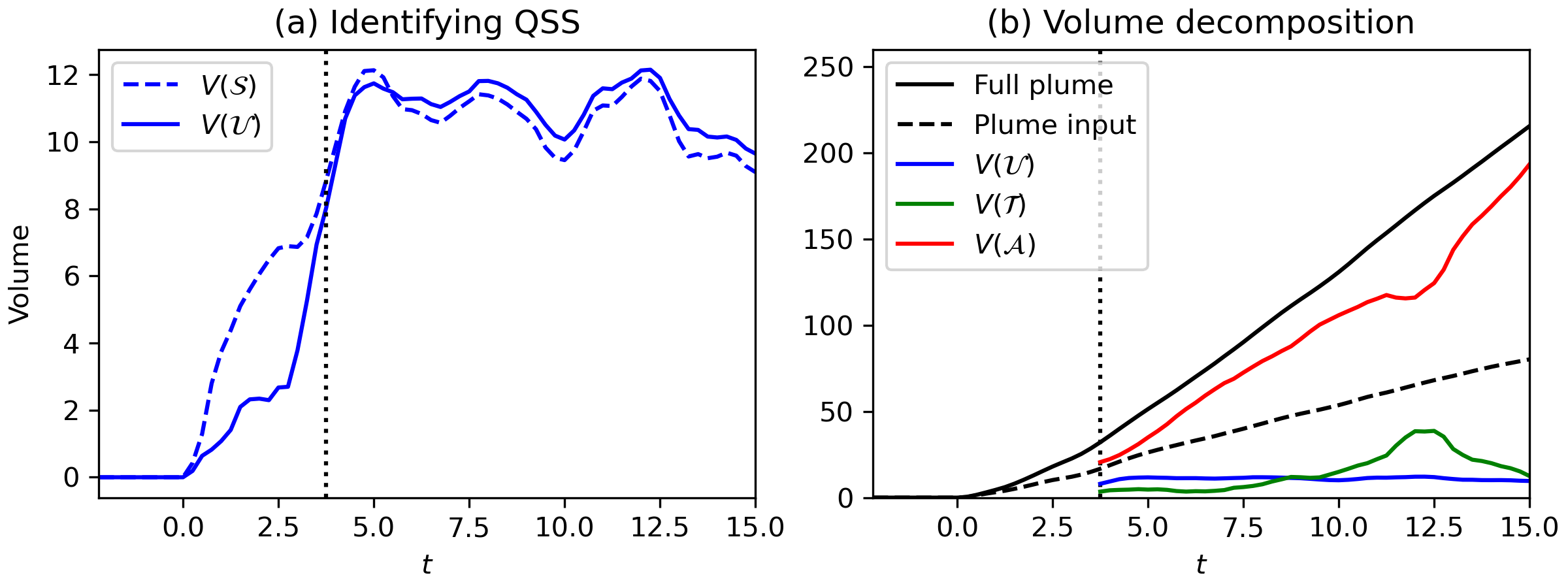}
	\caption{(a) Volume of the source line $\mathcal{S} = \{ \int_0^t S(b, \phi; t') \, \mathrm{d}t' > 0\}$ 
 (dashed line) and the region $\mathcal{U} = \{ M(b, \phi; t) < 0\}$ (solid line). The start of quasi-steady state (vertical dotted line)
 is identified as the time when these volumes agree to within $10\%$. (b) Decomposition of the full plume volume 
 into undiluted plume fluid, plume cap, and intrusion (solid coloured lines) using the partitioning introduced in 
 \S\ref{sec:partitioning}. The total plume input volume up to time $t$ (black dashed line) is shown for reference.}
	\label{fig:qss}
\end{figure}

To quantitatively identify the time at which quasi-steady state (QSS) starts, first we define the source line more 
generally as the region where the cumulative source volume is positive, $\mathcal{S} = \{ (b, \phi) \vert \int_0^t 
S(b, \phi; t') \,\mathrm{d}t' > 0\}$. Next we can define the volume associated with any region $\mathcal{R}$ of 
buoyancy-tracer space at time $t$ as
\begin{equation}
	V(\mathcal{R}; t) = \int_{\mathcal{R}} W(b, \phi; t) \, \mathrm{d}b \mathrm{d}\phi = \sum_\mathcal{R}
	W_{ij} (t).
\end{equation}
We expect that in quasi-steady state $\mathrm{d}V(\mathcal{S})/\mathrm{d}t \approx 0$. However, `plume bobbing' (i.e. 
the up-and-down motion of the plume cap noted in \S\ref{sec:flow}) results in some variation of the volume of the 
source line. The quasi-steady volume of undiluted plume fluid $V(\mathcal{S})$ can also gradually increase over time 
owing to the shielding effect. We therefore introduce an alternative constraint for identifying QSS which utilises the 
net mixing effect distribution $M$. As illustrated in figure~\ref{fig:VDcorr}, the source line represents undiluted 
plume fluid arriving in the stratified layer, which introduces volume into the distribution that is eventually mixed 
away from the source line. In the transient penetration stage, turbulent mixing redistributes fluid on the source line 
before mixing with the environment. Hence there is some accumulation on parts of the source line and $M>0$. However, once the plume reaches QSS and mixing with the environment continuously removes volume from the 
source line, $M$ must become negative. Away from the source line, $S$ vanishes so $M$ is necessarily 
positive according to \eqref{def:M}. We define the region $\mathcal{U} = \{ (b, \phi) \vert M(b, \phi; t) < 0\}$ and 
identify QSS as the time when the volume associated with $\mathcal{U}$, $V(\mathcal{U})$, is within $10\%$ of the 
volume of the source line $V(\mathcal{S})$. These volumes and the time we identify as the start of QSS, 
$t_{\mathrm{QSS}} \approx 3.5$, are shown in figure~\ref{fig:qss}(a). 

\subsection{Source, transport and accumulation regions}
\label{sec:partitioning}

We now restrict attention to quasi-steady state (QSS) $t > t_{\mathrm{QSS}}$, i.e. ignoring any transient dynamics during initial 
penetration. Here, we discuss the results for $M$ and show that this distribution can be used to partition plume fluid 
into three classes which identify coherent regions of the plume.

\begin{figure}
	\centering
	\includegraphics[width=\textwidth]{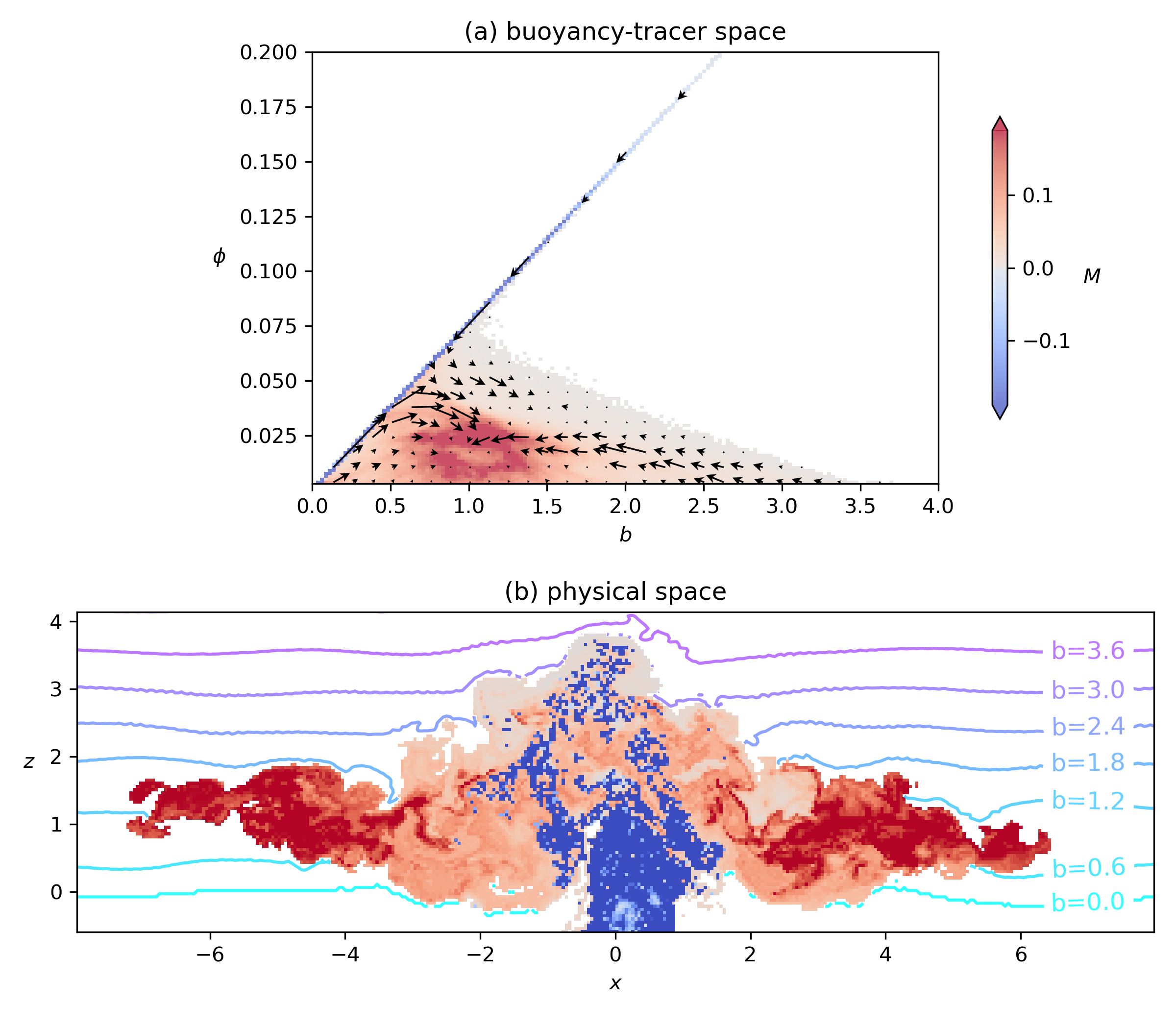}
	\caption{Snapshots at non-dimensional time $t=14$ of the buoyancy-tracer net mixing effect
	distribution, $M(b, \phi; t)$, in buoyancy-tracer space (top) and (bottom) in an $x-z$ cross-section of physical space. Buoyancy 
        contours are shown in the surrounding environment. The mixing flux distribution, $\boldsymbol{F}$, is overlaid in the top panel and the vectors are uniformly scaled to be visible.}
	\label{fig:Mplot}
\end{figure}

Figure~\ref{fig:Mplot} shows the net mixing effect distribution in both physical and buoyancy-tracer
space at $t = 14$, with the mixing flux distribution vectors $\boldsymbol{F}$ overlaid in buoyancy-tracer space. 
The distribution is represented in physical space by using the buoyancy and tracer concentration 
to map between buoyancy-tracer space and physical space, i.e. we plot $M(b(\boldsymbol{x},t), \phi(\boldsymbol{x},t); 
t)$. To avoid confusion between regions of buoyancy-tracer space and the
corresponding fluid in physical space, we refer to the former as regions and the latter as classes of fluid.
Recall that the net mixing effect distribution, $M$, quantifies the integrated effect of the mixing flux distribution, 
$\boldsymbol{F}$, or, equivalently, the volume difference between the volume distribution and the cumulative source 
distribution, representing the volume change due to mixing. As expected, we find $M < 0$ on the source line where 
undiluted plume fluid is continuously supplied before being mixed away into the $M > 0$ region. Environmental fluid is 
entrained into the plume via mixing and accumulates in the intrusion where $M$ is maximised. We define \emph{class U} 
as undiluted plume fluid corresponding to the \emph{source region} $\mathcal{U} = \{ M < 0 \}$, introduced in the 
definition of QSS in \S\ref{sec:qss}. For convenience, we use the notation $\{ M < 0 \}$ as shorthand for $\{ (b, 
\phi) \vert M(b, \phi; t) < 0\}$ henceforth. The mixing flux vectors point 
along the source line, indicating that mixing within $\mathcal{U}$ is mostly within the plume rather than between the 
plume and environment, owing to the shielding effect. Once undiluted plume fluid is exposed to the environment, there 
is a strong mixing flux between $\mathcal{U}$ and the $\phi=\phi_{\min}$ axis where environmental fluid joins the 
distribution.

As discussed in \S\ref{sec:defn}, we expect volume to accumulate in some region of buoyancy-tracer space. This is 
clearly demonstrated in physical space, where fluid collects in the intrusion after mixing with the environment.
We can distinguish the accumulation region from the `transport' region through which volume moves 
to reach the accumulation region by identifying a region in which $M$ is small and approximately constant. In this region, plume fluid is actively mixing with the environment and transporting volume away from the source line, but fluid
does not accumulate in this region. Then, fluid that has undergone significant mixing and accumulates in the intrusion corresponds 
with a region where $M$ is increasing with time. The regions are distinguished by a time-dependent threshold $m^*(t)$ 
such that the \emph{transport region}, where much of the transition from undiluted to mixed plume fluid occurs, is 
$\mathcal{T} = \{0 < M \le m^*\}$ and we refer to fluid corresponding to this region as \emph{class T}. The 
\emph{accumulation region} is $\mathcal{A} = \{M > m^*\}$ with corresponding mixed fluid accumulating in the intrusion 
referred to as \emph{class A}. The threshold $m^*(t)$ is found by identifying the value $m$ which 
minimises the total mixing flux convergence $f(m; t)$ associated with a region $\{ 0 < M \le m(t)\}$ at each time $t$, 
where
\begin{equation}
	f(m; t) = \int_{\{ 0 < M \le m \}} \frac{\mathrm{d} W}{\mathrm{d} t} - S \, \mathrm{d}b\mathrm{d}\phi =
	-\int_{\{0 < M \le m\}} \bnabla_{(b,\phi)} \cdot \boldsymbol{F} \, \mathrm{d}b\mathrm{d}\phi, \label{eq:conv}
\end{equation}
so that the volume that enters the region $\mathcal{T}$ is approximately equal to the volume leaving
$\mathcal{T}$. Then, remaining mixed fluid lies in the region $\mathcal{A}$ which must capture the 
accumulation of mixed fluid, formed from undiluted plume fluid that has entered the stratified layer and entrained 
environmental fluid. The numerical implementation of this method using the discrete form of the volume distributions 
is discussed further in appendix~\ref{app:D}. The time variation of the threshold $m^*(t)$ is shown in figure~\ref{fig:m_thresh_calc} in appendix~\ref{app:D}.

\begin{figure}
	\centering
	\includegraphics[width=\textwidth]{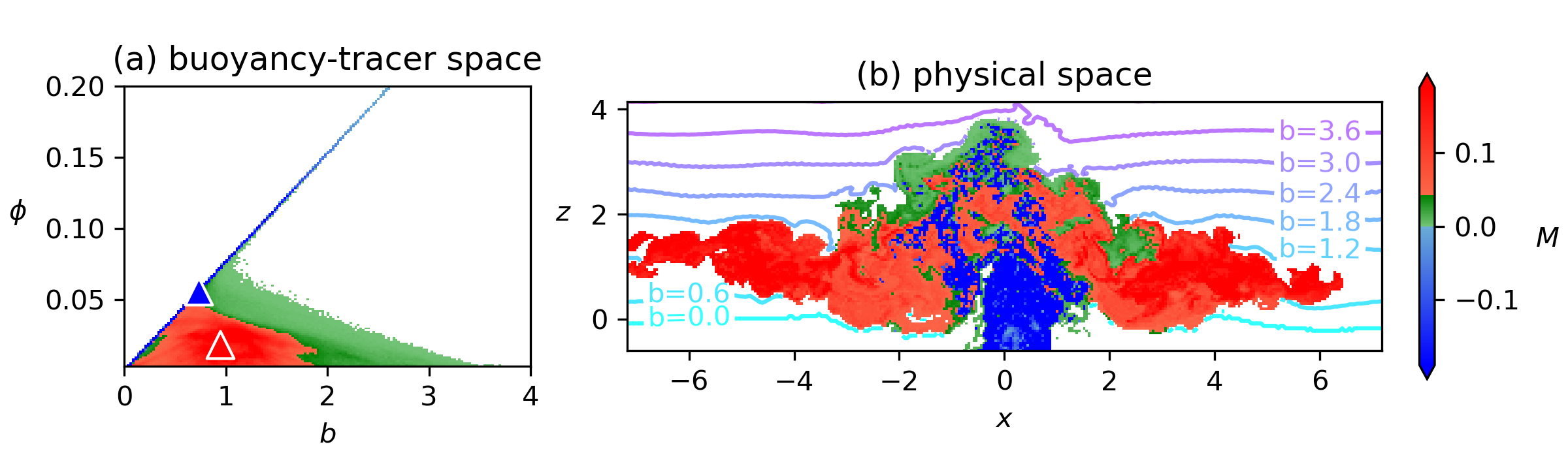}
	\caption{As in figure~\ref{fig:Mplot}, with the distribution partitioned into three regions: $\mathcal{U}$ (blue) where $M \le 0$, $\mathcal{T}$ (green) where $0 < M \le m^*(t)$ and $\mathcal{A}$ (red) where $M > m^*(t)$. The threshold $m*(t)$ minimises the total mixing flux convergence defined by \eqref{eq:conv}. Corresponding fluid classes U, T and A (respectively) shown in physical space. Triangles represent the volume-weighted centre-of-mass in class U (blue) and class A (red). Undiluted plume fluid in $\mathcal{U}$ is mixed towards the blue triangle whilst fluid accumulating in $\mathcal{A}$ is mixed towards the red triangle.}
	\label{fig:M_partitioned}
\end{figure}

The net mixing effect distribution with this partitioning is shown in figure~\ref{fig:M_partitioned},
with class U coloured blue, class T coloured green and class A coloured red. Within classes U and A, the volume
weighted centre-of-mass is shown by a coloured triangle, approximately indicating the position in buoyancy-tracer 
space towards which mixing acts to move fluid within each class. Internal plume mixing of undiluted plume fluid 
redistributes volume on the source line towards the blue triangle and homogenisation of mixed fluid in the intrusion 
accumulates volume near the red triangle. Figure~\ref{fig:M_partitioned} demonstrates the correspondence between the 
regions $\mathcal{U}$, $\mathcal{T}$ and $\mathcal{A}$ and coherent structures of the plume. The source region 
$\mathcal{U}$ identifies the rising undiluted plume. The accumulation region $\mathcal{A}$ identifies the radially 
spreading intrusion. The transport region $\mathcal{T}$ corresponds with newly-generated mixtures in the plume cap and 
subsiding fluid joining the intrusion.  The partition of the full plume volume into the undiluted plume, plume cap, 
and intrusion volume is shown in figure~\ref{fig:qss}(b) for $t > t_{\mathrm{QSS}}$. As expected, we find that the 
volume of the intrusion (class A) dominates the plume at late times, as the volume of undiluted plume fluid and the 
plume cap each remain approximately constant.

\subsection{Entrainment}
\label{sec:entrainment}
We calculate the entrained volume $E(t)$ as the difference between the volume of the full
plume and the cumulative volume of the source term $S$ up to time $t$, so that $E$ represents the volume of
environmental fluid that has been mixed into the plume up to time $t$. We have
\begin{align}
	E(t) &\equiv \int_{\{W > 0\}} W(b, \phi; t) \, \mathrm{d}b \mathrm{d}\phi - \int_0^t\int_{\{W > 0\}} S(b,
	\phi; t') \, \mathrm{d}b \mathrm{d}\phi \mathrm{d}t' \\ 
	  &= -\int_0^t \int_{\{W>0\}} \nabla_{(b, \phi)} \cdot \boldsymbol{F}(b, \phi; t') \, \mathrm{d}b \mathrm{d}\phi
	  \mathrm{d}t' \\
	  &= \int_0^t \int_{\{\phi=\phi_{\min}\}} F_{\phi}(b, 0; t') \,\mathrm{d}b \mathrm{d}t' 
	  \approx \sum_{t'}\sum_{i} \frac{F^\phi_{i,j=0}(t')}{\Delta \phi} \Delta t', \label{eq:ent}
\end{align}
where the second equality follows from time-integrating \eqref{eq:VDbudget} and the final equality follows from
Green's theorem and the fact that the mixing flux $\boldsymbol{F}$ vanishes on the boundary of the $\{W>0\}$
region except on the surface $\phi = \phi_{\min}$ where environmental fluid enters the volume distribution via
entrainment. The numerical calculation of $E$ using the discrete form of the mixing flux distribution $F_{ij}^{\phi}$ 
is given by \eqref{eq:ent}.

\begin{figure}
	\centering
	\includegraphics[width=\textwidth]{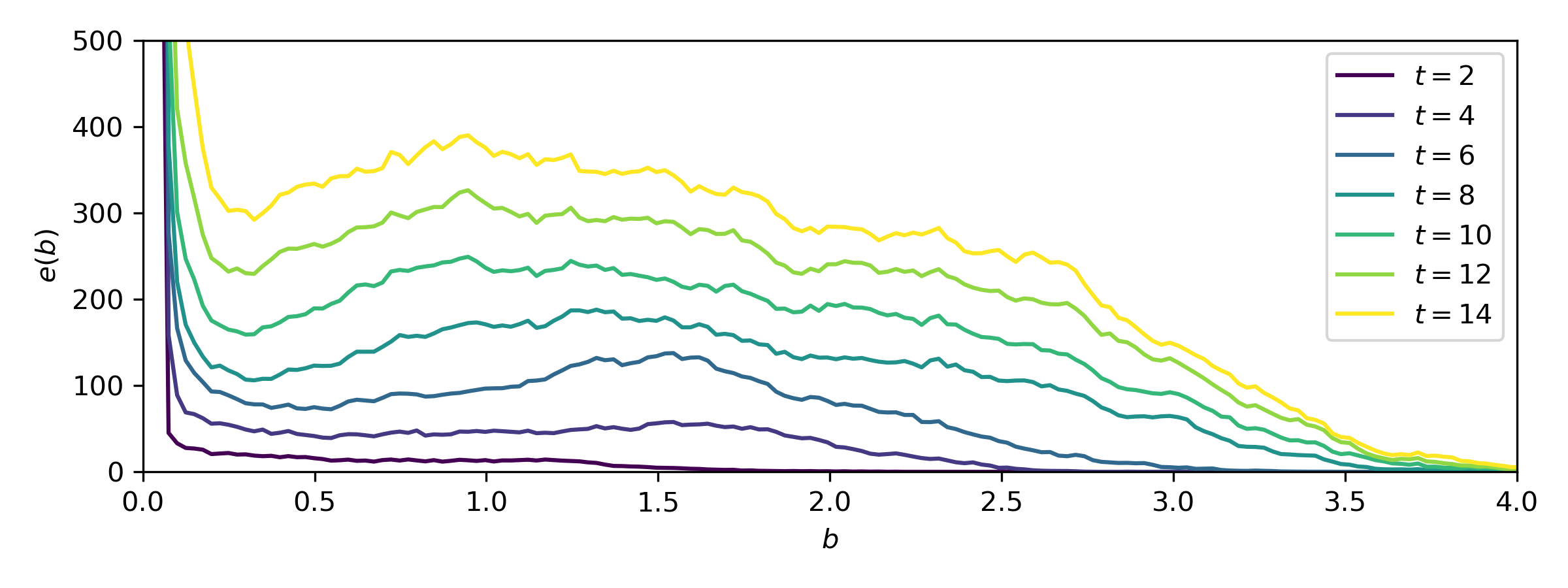}
	\caption{Entrainment profile $e(b, t)$ defined by \eqref{eq:ent_profile} at fixed time intervals 
 post-penetration. The value $e(b,t)$ is the volume of environmental fluid entrained up to time $t$, per unit 
 buoyancy, as a function of buoyancy $b$.}
	\label{fig:ent}
\end{figure}

A common definition of an `entrainment profile' with respect to height is the fractional volume (or mass) increase 
with height (e.g. \citet{derooy2013}). This is not a useful definition in the case where the plume overturns, since 
the rate of change with height captures multiple components of the plume which act to entrain fluid in (potentially) 
different ways. Exploiting the linear increase of buoyancy with height in the initially linearly stratified 
environment, we treat buoyancy as a rough proxy for height and define an entrainment profile with respect to buoyancy,
\begin{equation}
    e(b_i, t) = \int_0^t F_\phi(b_i, 0; t') \, \mathrm{d}t' \approx \sum_{t'} \frac{F_{i,j=0}^\phi(t')}{\Delta \phi \Delta b} \Delta t', \label{eq:ent_profile}
\end{equation} 
which represents the volume of environmental fluid entrained up to time $t$ per unit buoyancy. We can then 
define the volume entrained into a region $\mathcal{R}$ of buoyancy-tracer space which intersects the $\phi = 
\phi_{\min}$ boundary as
\begin{equation}
    E(\mathcal{R}, t) = \int_0^t\int_{\{\phi=\phi_{\min}\}\cap\mathcal{R}} F_\phi(b, 0; t') \, \mathrm{d}b \mathrm{d}t' = \int_{\{\phi=\phi_{\min}\}\cap\mathcal{R}} e(b, t) \, \mathrm{d}b, \label{eq:entregion}
\end{equation}
which is numerically calculated by summing over the indices $i$ in \eqref{eq:ent_profile} which belong to the 
intersection of $\mathcal{R}$ with the boundary $\phi=\phi_{\min}$. We define the entrainment rate as the time rate 
of change of the entrained volume into a region $\mathcal{R}$, i.e. $\dot{E}(\mathcal{R}) = \partial_t 
E(\mathcal{R})$. This quantifies the rate at which volume is entrained into a physical volume represented by a 
region $\mathcal{R}$ in buoyancy-tracer space. Whilst we expect vigorous mixing at the boundary between a 
sub-volume of the plume and the environment to result in entrainment, the entrainment rate $\dot{E}$ does not 
necessarily quantify this since larger volumes would be expected to entrain more volume over time even if the 
`strength' of the mixing is weaker. To quantify the strength of the entrainment into each sub-volume of physical space 
corresponding to a partitioning of buoyancy-tracer space, we define the \emph{specific entrainment rate} as the ratio 
of the entrainment rate with the volume of each sub-region itself, i.e. $\dot{E}(\mathcal{R}) / V(\mathcal{R})$ for 
each region $\mathcal{R} = \mathcal{U}, \mathcal{T}, \mathcal{A}$.

\begin{figure}
	\centering
	\includegraphics[width=\textwidth]{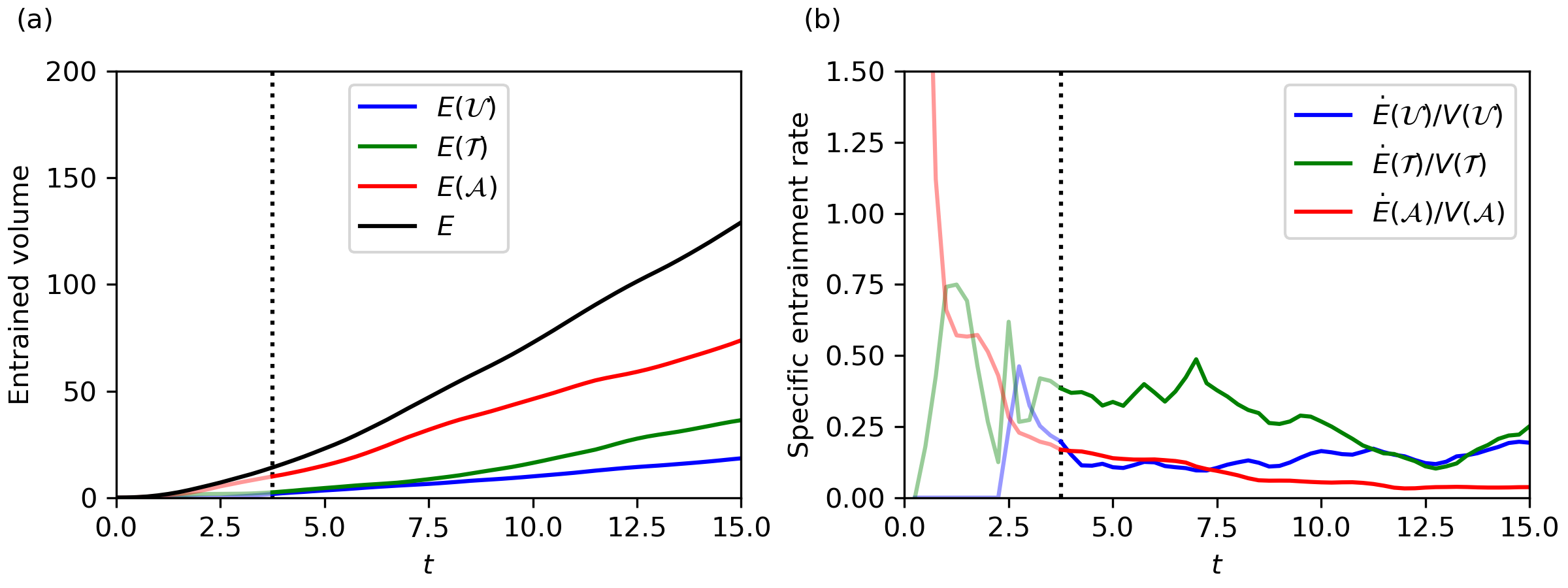}
	\caption{(a) Volume of environmental fluid entrained up to time $t$ (black line) and its decomposition into 
 entrainment into the regions $\mathcal{U}, \mathcal{T}$ and $\mathcal{A}$ (coloured lines). These regions are 
 defined in \S\ref{sec:partitioning} and the entrained volumes $E(\mathcal{U}), E(\mathcal{T})$ and $E(\mathcal{A})$ 
 are computed using \eqref{eq:entregion}. (b) Specific entrainment rate in each class, defined as the ratio of the 
 rate of change of the entrained volume and the volume, e.g. $\dot{E}(\mathcal{U})/V(\mathcal{U})$. In both plots, the vertical dotted line indicates the start of the quasi-steady state.}
	\label{fig:entrainment}
\end{figure}

Figure~\ref{fig:ent} shows the evolution of the entrainment profile through the simulation. Significant
entrainment occurs over a large range of intermediate buoyancy values, indicating that most entrained volume is from 
the environmental fluid surrounding the intrusion. Using the partitioning introduced in \S\ref{sec:partitioning}, the 
entrained volume can be decomposed into the volume entrained into the plume cap $E(\mathcal{T})$, the volume 
entrained into the intrusion $E(\mathcal{A})$, and the volume entrained just before penetrating the stratified 
layer $E(\mathcal{U})$. Figure~\ref{fig:entrainment}(a) shows the entrained volume as well its decomposition. 
Here, a correction has been made to $E(t)$ to account for numerical artefacts -- see appendix~\ref{app:B} for details. 
Figure~\ref{fig:entrainment}(b) shows the entrainment rate in each of the 
regions $\mathcal{U}, \mathcal{A}$ and $\mathcal{T}$. The relative contribution of plume cap entrainment 
and `lateral' entrainment in the intrusion to the overall entrainment is noted as an open
question in the study of fountains by \citet{hunt2015}, where the plume cap is analogous to the `fountain-top'. 
Here we find that the contribution to the entrained volume from the plume cap (class T) is weak compared with 
the intrusion (class A) when in quasi-steady state. Under the definition of quasi-steady state for this flow given in 
section~\ref{sec:qss}, volume continuously accumulates in the intrusion and hence the volume of the intrusion dominates
the volume of the full plume at late times. Since the intrusion spreads radially, there is a greater contact area 
between the intrusion and the surrounding environmental fluid compared with the plume cap and the rising plume. This 
allows a greater volume of environmental fluid to be entrained into the intrusion. This suggests that entrainment of 
environmental fluid from the lower part of the stratified layer into the intrusion is important for setting the centre-of-
mass of the quasi-steady state buoyancy-tracer distribution. However, mixing during the overturning process in the plume 
cap near $\zmax$ is important for setting the maximum accessible buoyancy of the volume distribution after mixing, and 
therefore the extent of the accumulation region in buoyancy-tracer space. The specific entrainment rate is larger in class 
T than in class A, indicating stronger mixing with the environment in the plume cap in quasi-steady state compared with the 
intrusion.

\subsection{Three-stage mixing process}
\label{sec:stages}

\begin{figure}
	\centering
	\includegraphics[width=\textwidth]{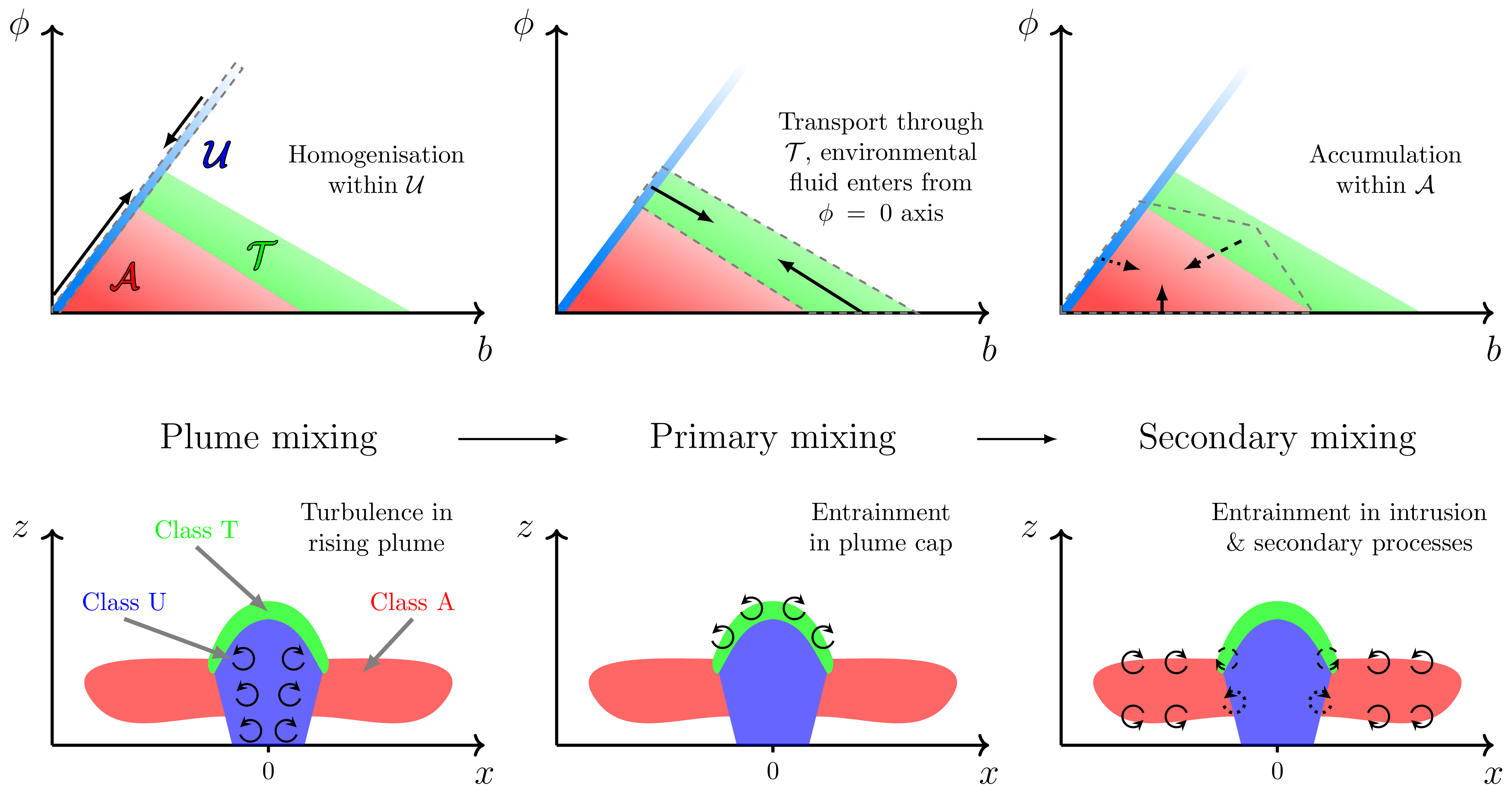}
	\caption{Schematic of the three stage mixing process in quasi-steady state convective penetration of a 
 buoyant plume into a stably stratified layer, identified by the partitioned buoyancy-tracer volume distribution. 
 Buoyancy-tracer space is shown on the top row, with arrows indicating the movement of volume within each mixing 
 stage. The region of buoyancy-tracer space affected by mixing in each stage is indicated by a gray dashed envelope. 
 Physical space is shown on the bottom row with circular arrows indicating where mixing is located. In the secondary 
 mixing stage the three distinct mixing processes are shown as dotted, dashed and solid arrows, both in physical and 
 buoyancy-tracer space.}
	\label{fig:mixing_schematic}
\end{figure}

Together, the results presented in this section suggest a multi-stage mixing process in 
quasi-steady state convective penetration. This is summarised in figure~\ref{fig:mixing_schematic}, in which we show 
schematics of the volume distribution partitioned into three regions $\mathcal{U}, \mathcal{T}$, and $\mathcal{A}$ 
of buoyancy-tracer space, and the corresponding classes of fluid in cross-sections of the plume. Fluid moves through
each stage of mixing from left to right, corresponding with an increasing value of the net mixing effect $M$, though 
occasionally the primary mixing stage may be skipped -- see following discussion. In each stage, the convex envelope 
of fluid involved in mixing is indicated by a gray dashed line in $(b,\phi)$-space; arrows in $(b,\phi)$-space 
indicate the movement of individual fluid parcels due to the mixing process; and circular arrows in the $x-z$ cross-
section indicate the \emph{location} of mixing in physical space. These circular arrows are illustrative and are not 
intended to indicate the physical nature of the mixing mechanism in each stage. In the secondary mixing stage where 
multiple mixing processes occur simultaneously, three arrow types are shown which correspond between physical and buoyancy-
tracer space. The mixing stages in QSS convective penetration are as follows:

\begin{enumerate}
	\item Mixing within the rising plume as undiluted plume fluid penetrates into the stratified layer. This fluid
            is shielded from the surrounding environment. Thus, in buoyancy-tracer space, internal mixing of undiluted 
            plume fluid acts within $\mathcal{U}$ only, homogenising the distribution and consequently moving fluid 
            towards the CoM of class U indicated by the blue triangle in figure~\ref{fig:M_partitioned}. This sets the
		buoyancy-tracer characteristics of plume fluid that is first exposed to environmental fluid near
		$z_{\max}$. 

	\item Primary mixing between the plume and environment occurs in the plume cap. This mixing may extend below 
            the plume cap into the intrusion as	the mixing timescale is slow compared to the dynamical timescale. In 
            buoyancy-tracer space, the primary mixing stage acts on the undiluted plume fluid 
            concentrated near the class U CoM and the surrounding environmental fluid with values of buoyancy close to 
            $b = z_{\max}$. The resulting mixtures are transported towards intermediate values of 
            buoyancy as indicated by the mixing flux distribution in figure~\ref{fig:Mplot}(a).

	\item After primary mixing with the environmental fluid, subsiding mixed fluid joins the intrusion and     
            homogenises with its surroundings, moving from class T to class A (dashed arrows). A 
            number of mixing processes act on the fluid accumulating in class A. In buoyancy-tracer space, mixing 
            occurs between four regions of fluid as indicated by the convex envelope in 
            figure~\ref{fig:mixing_schematic}. Secondary mixing with the environment occurs as environmental fluid 
            lower in the stratified layer is entrained into the intrusion as it spreads (solid arrows), dominating the 
            volume entrained by the full plume at late times. At the edges of the rising plume, undiluted plume 
            fluid mixes with fluid already in the intrusion, resulting in some fluid parcels moving directly from 
            class U to class A without entering the primary mixing stage (dotted arrows) as seen in 
            figure~\ref{fig:Mplot}(a). Finally, mixing in the interior of the intrusion 
            homogenises the buoyancy-tracer distribution of fluid that accumulates within region $\mathcal{A}$. 
            As fluid moves radially, large volumes of fluid in class A are concentrated near the CoM 
            in $(b,\phi)$-space, where $M$ (and $W$) are largest.
\end{enumerate}

%%%%%%%%%%%%%%%%%%%%%%%%%%%%%%%%%%%%%%%%%%%%%%%%%%%%%%%%%%%%%%%%%%%%%%%%%%%%%%%%%%%%%%%%%%%%%%%%%%%%%%%%%%%%%

\section{Mixing diagnostics}

In this section we use the partitioning introduced in \S4 to examine the statistics of mixing  in each stage of the plume 
evolution.

\subsection{Characterising mixing}
\label{sec:mixvars}

To characterise the physical nature of the mixing in each stage, we consider the mixing efficiency which relates the total energy expended in turbulent mixing with the actual mixing
achieved (e.g. \citet{wykes2015}). The most useful definition depends on context, e.g. \citet{gregg2018} for
oceanographic contexts and \citet{chemel2007} for an atmospheric setting. In buoyancy-driven stratified flows, the 
mixing efficiency is quantified by utilising the partitioning of potential energy into available potential energy 
(APE) and background potential energy (BPE). For an incompressible, Boussinesq flow BPE is the potential energy that 
is not available to do work whilst APE represents the energy stored in the buoyancy field if the flow is not in 
gravitational equilibrium \citep{wykes2015, lorenz1955}. Turbulent mixing irreversibly converts APE into BPE and 
results in dissipation of turbulent kinetic energy (TKE). The conversion of APE into BPE quantifies the energy 
expenditure that results in mixing of the buoyancy field and its sum with the TKE dissipation represents the total 
energy expended by turbulent mixing. The ratio of these two quantities forms the mixing efficiency. Following 
\citet{howland2020} and \citet{holliday1981}, in the case where $\partial_z b$ is constant in the initial stratified 
environment we may treat the quadratic form of the potential energy,
\begin{equation}
	E_p = \frac{1}{2} \langle b'^2 \rangle,
\end{equation}
as a proxy for available potential energy. Here, $b'(\boldsymbol{x},t) = b(\boldsymbol{x},t) - z$ is the departure 
from the linear initial stratification and $\langle \cdot \rangle$ denotes a volume average. We refer to $E_p$ as 
the perturbation potential energy (PE). A full derivation of the perturbation PE budget is given in 
appendix~\ref{app:E}, which follows the derivation as described in \citet{howland2020} except with 
SGS terms included. The irreversible conversion of perturbation PE to BPE that results from the 
reduction of buoyancy gradients by mixing is captured by the buoyancy variance dissipation rate,
\begin{equation}
	\overline{\chi} = \left\langle \kappatot \,|\bnabla b' |^2 \right\rangle, \label{eq:chi}
\end{equation}
where $\kappatot = (\Rey \Pr)^{-1} + \kappaSGS^{(b)}$ is the total diffusivity of buoyancy. The buoyancy variance 
dissipation rate represents the primary sink of $E_p$. The total energy 
dissipated via turbulent mixing is the sum of $\overline{\chi}$ and the dissipation rate of TKE,
\begin{equation}
	\overline{\varepsilon} = \left\langle \nutot	\frac{\partial u_i}{\partial x_j}
	\frac{\partial u_i}{\partial x_j} \right\rangle, \label{eq:tked}
\end{equation}
where $\nutot = \Rey^{-1} + \nuSGS$ is the total viscosity. The TKE dissipation rate acts as the 
primary sink of turbulent kinetic energy. The instantaneous mixing efficiency $\eta$ is then
\begin{equation}
	\eta = \frac{\overline{\chi}}{\overline{\chi}+\overline{\varepsilon}},
\end{equation}
as in e.g. \citet{howland2020, peltier2003}. Here, we use an overbar to denote a volume averaged quantity. It is instructive to examine the spatial structure of the dissipation rates. In particular, 
regions of large  $\chi$ indicate intense buoyancy gradients and regions of large $\varepsilon$ indicate intense 
turbulent motion.

Further information on the state of turbulence in stratified flows is often drawn from the buoyancy Reynolds number 
$\Reb \equiv \langle \varepsilon \rangle / \nu N^2$. We define a pointwise activity parameter $I$ and its mean 
$\overline{I}$,
\begin{equation}
	I = \frac{\frac{\partial u_i}{\partial x_j} \frac{\partial u_i}{\partial x_j}}{\left|\frac{\partial
	b}{\partial z}\right|} = \frac{\varepsilon}{\nutot \left|\frac{\partial b}
        {\partial z}\right|}, 
        \quad
        \overline{I} = \frac{\langle \varepsilon /\nutot \rangle}{\langle \left| 
                        \frac{\partial b}{\partial z} \right| \rangle}. \label{eq:I}
\end{equation}
The bulk property $\overline{I}$ is analogous to $\Reb$ except with SGS contributions to viscosity included. Also, 
we replace the global buoyancy timescale $N^{-1}$ with a local measure of the buoyancy timescale $\lvert\partial_z 
b\rvert^{-1}$ given by the local buoyancy gradient. This is a more appropriate measure since buoyancy gradients within 
the plume differ significantly from the background linear stratification and are more representative of the regime in 
which mixing occurs inside the plume. As with $\Reb$, the mean activity parameter $\overline{I}$ can be interpreted as the ratio 
of the destabilising effects of turbulent stirring to the stabilising effects of buoyancy and viscosity. Similarly, 
$I$ may be treated as the ratio of the (local) buoyancy timescale $(\partial b / \partial z)^{-1/2}$ to the timescale 
of development of turbulent effects $\left(\varepsilon / \nutot\right)^{-1/2}$ 
\citep{ivey2008}. Regions of large $I$ indicate active turbulence \citep{garcia2011} that is weakly affected by 
stratification. 

\begin{figure}
	\centering
	\includegraphics[width=\textwidth]{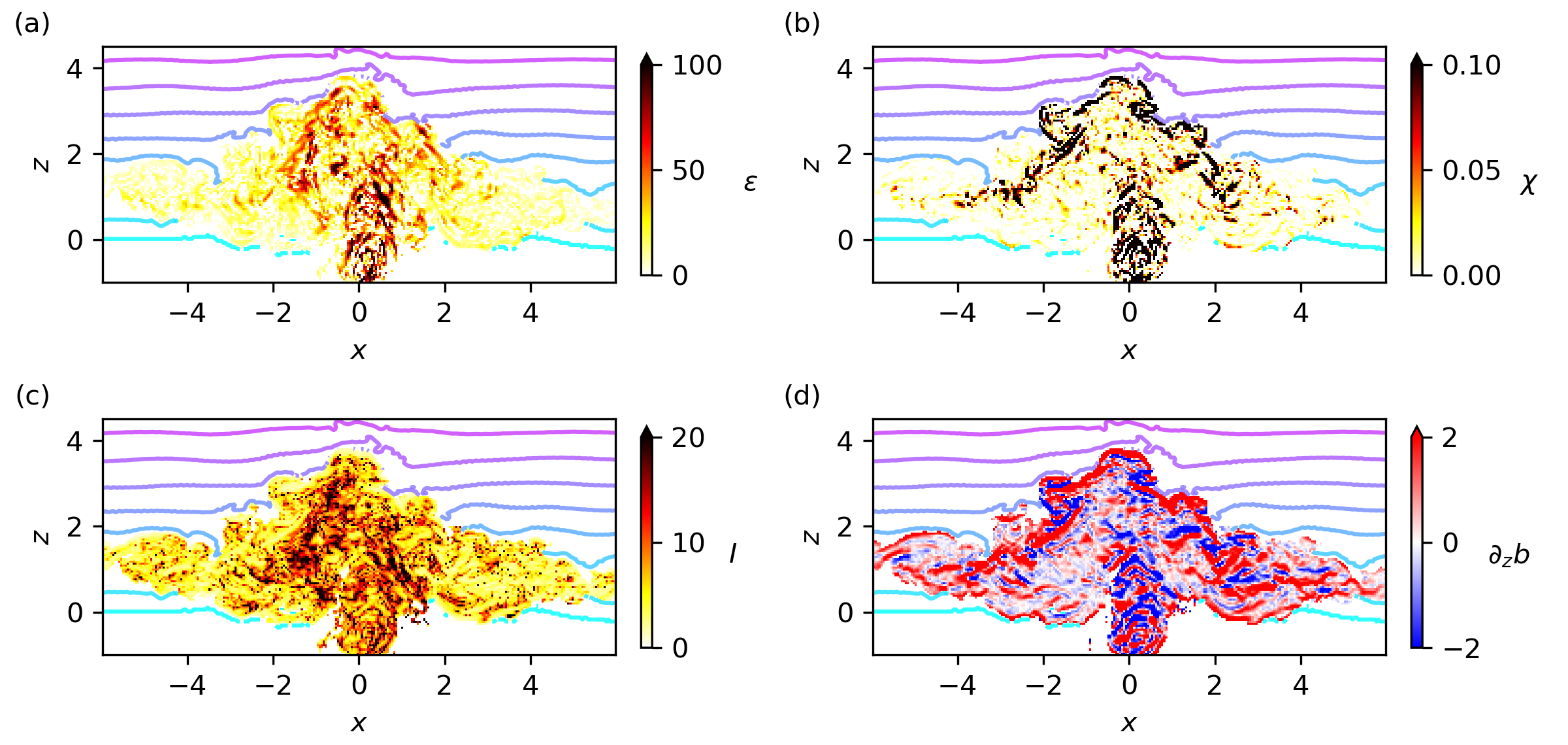}
	\caption{$x-z$ cross sections of the mixing diagnostics $\varepsilon, \chi, I$ and $\partial_z b$ within the 
 plume, where $\phi > \phi_{\min}$ at non-dimensional time $t=14$. Mixing diagnostics are defined in 
 \S\ref{sec:mixvars}. Cross-sections are taken at the plume centreline. Buoyancy contours are shown outside the plume.}
	\label{fig:cross_sections}
\end{figure}

\subsection{Results \& discussion}

\begin{figure}
	\centering
	\includegraphics[width=\textwidth]{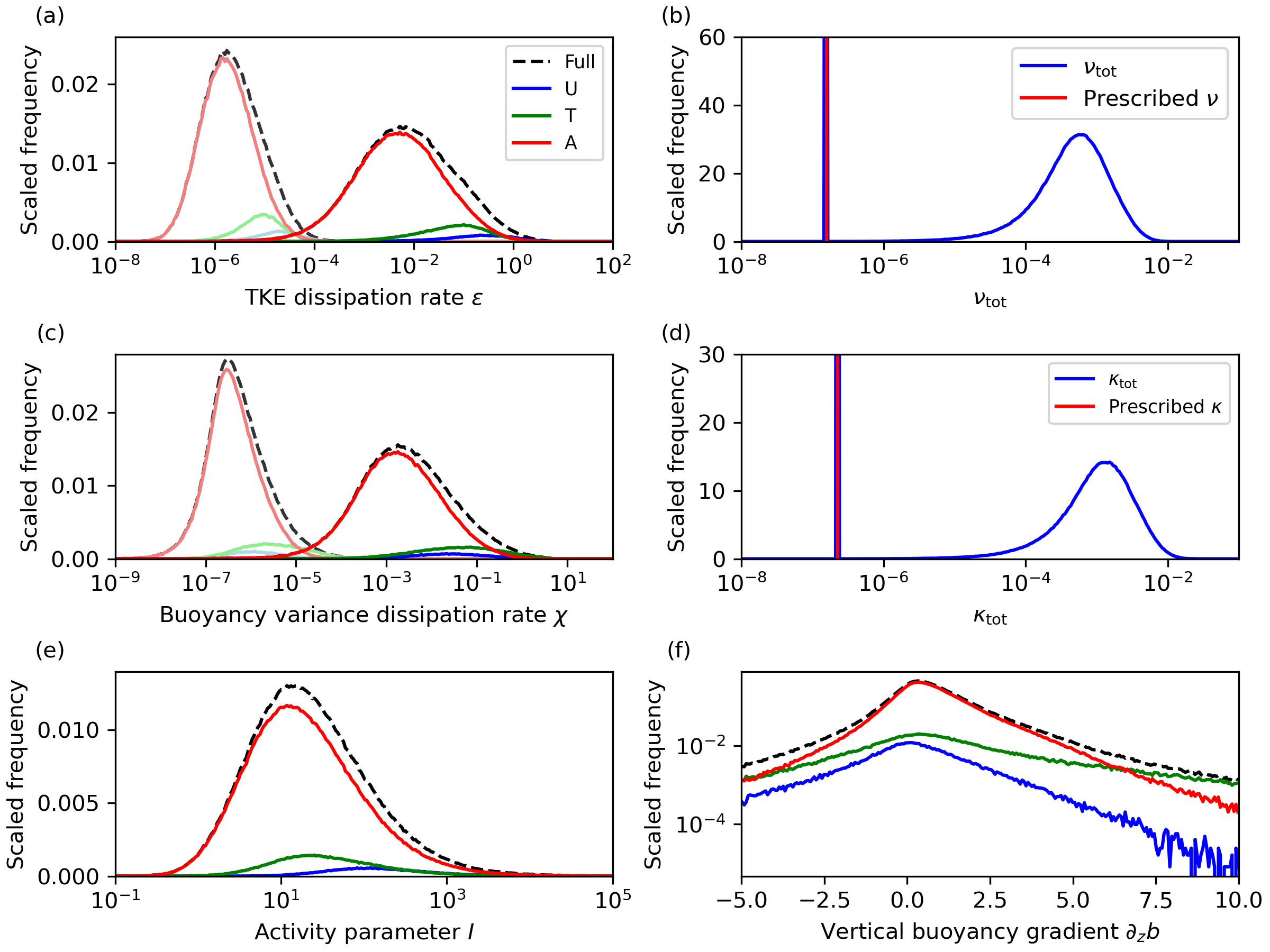}
	\caption{(a), (c), (e), (f) show histograms of the mixing diagnostics $\varepsilon, \chi, I$ and $\partial_z 
 b$ defined in \S\ref{sec:mixvars} (black dashed line) decomposed into histograms within class U, T, A (coloured 
 lines). The histograms within each class are scaled so that their sum is the histogram of the full plume. In (a) 
 and (c), the histograms are further separated into those with $\nuSGS=0$ and $\kappaSGS^{(b)}=0$ (light colours) 
 and with $\nuSGS$ and $\kappaSGS$ non-zero (usual colours). (b) and (d) show histograms of the total viscosity $\nutot$ and total buoyancy 
 diffusivity $\kappatot$ respectively. All histograms are computed at $t=15$. All vertical axes show the 
 scaled histogram frequency. The scaling is such that the total area is unity for the full plume histograms and the 
 sum of the partitioned histograms equals the full plume histogram. Similarly in (b) and (d) the frequency is scaled 
 so that the total area is unity.}
	\label{fig:pdfs}
\end{figure}

Cross-sections in 
physical space of the dissipation rates $\varepsilon$ and $\chi$, activity parameter $I$ and local buoyancy gradient 
$\partial_z b$ are shown in figure~\ref{fig:cross_sections}. These diagnostics are only shown in the plume, where 
$\phi > \phi_{\min}$. In the surrounding environment, buoyancy contours are shown. The 
mixing diagnostics within the plume structures identified by classes U, T and A are quantified by histograms 
separated into each class in figure~\ref{fig:pdfs}. The colours for each class correspond with those used in 
\S\ref{sec:VD} (e.g. figure~\ref{fig:mixing_schematic}). The black dashed lines show the histograms for the full
plume, i.e. all fluid within the stratified layer where $\phi > \phi_{\min}$. This full histogram is normalised to 
form a PDF. The partitioned histograms are scaled so that the sum of the class U, T and A histograms equals the 
full plume PDF. These histograms are summarised by volume averages of the mixing diagnostics within each class, as 
well as the full plume volume average, in table~\ref{tab:mixing}.

The histograms for TKE dissipation in figure~\ref{fig:pdfs}(a) and buoyancy variance dissipation in 
figure~\ref{fig:pdfs}(c) are further separated based on where $\nuSGS$ and $\kappaSGS$, respectively, are non-zero or 
vanish. This distinction is made since the total viscosity $\nutot$ and total dissipation $\kappatot$ are bimodal with 
a peak where the SGS contribution vanishes (where the simulation effectively switches from LES to direct numerical 
simulation, such that $\nutot$ and $\kappatot$ reduce to the molecular values) and a peak where the SGS contribution 
is non-zero -- see figures~\ref{fig:pdfs}(b) and (d). The separation of the histograms based on non-zero and zero SGS 
contribution demonstrates that the bimodality of the histograms for $\varepsilon$ and $\chi$
is a consequence of the bimodal total viscosity and diffusivity alone. The vertical buoyancy gradient histogram is 
shown on a logarithmic scale since the most extreme values are rare but remain important for mixing.

The results show that the mixing statistics are quantitatively different in each class, suggesting that the mixing 
regimes differ in each of the three stages of the plume evolution highlighted in \S\ref{sec:stages}. The 
cross-sections in figure~\ref{fig:cross_sections} show that TKE dissipation $\varepsilon$ is particularly large in the 
rising undiluted plume and in some regions of the plume cap. This is supported by the histograms which show that TKE 
dissipation is an order of magnitude larger in class U and T compared to class A. The activity parameter is 
also largest in the rising undiluted plume but comparable in the plume cap and intrusion. The largest values (both 
positive and negative) of the vertical buoyancy gradient $\partial_z b$ are found in the plume cap (class T) with a 
clear positive bias. The relatively tighter spread of the $\partial_z b$ distribution in class U in 
figure~\ref{fig:pdfs}(f) suggests that smaller vertical buoyancy gradients are more common in the undiluted plume 
fluid as compared with the plume cap and intrusion. This explains the increased magnitude of the activity parameter 
in class U compared with class T despite similar magnitudes of TKE dissipation; $\varepsilon$ is large in both 
classes but $\partial_z b$ is generally larger in class T and hence $I$, proportional to the ratio $\varepsilon / 
\partial_z b$, is smaller in class T. The strongest buoyancy gradients as measured by $\chi$ are found in the 
primary mixing region (class T), where fluid overturns and subsides fluid in the plume cap. Intense buoyancy 
gradients are also found in the rising column of undiluted plume fluid, but $\chi$ is more sparse here than in the 
plume cap. The strong buoyancy gradients found at the top edge of the plume cap are a result of the relatively less 
buoyant plume impinging on the more buoyant surrounding environmental fluid. Note that the largest values of $\chi$ 
at the extreme positive tail of the full plume PDF are almost all from class T.

\begin{table}
	\begin{center}
	\begin{filecontents*}{data.csv}
1.00e+02,3.64e+00,9.88e+00,8.65e+01
1.68e+01,8.16e+01,1.88e+01,1.28e+01
1.29e+00,1.41e+00,2.88e+00,1.11e+00
4.66e-02,3.47e-01,1.27e-01,2.47e-02
2.38e-02,5.95e-02,1.29e-01,1.03e-02
3.39e-01,1.47e-01,5.04e-01,2.94e-01
\end{filecontents*}
\pgfplotstabletypeset[col sep=comma, 
		create on use/new col/.style={create col/set list={
		Volume \%,
		Activity parameter \, $\overline{I}$,
		Vertical buoyancy gradient\, $\langle\left|\partial_z b\right|\rangle$,
		TKE dissipation rate\, $\overline{\varepsilon}$,
		Buoyancy variance dissipation rate\,$\overline{\chi}$,
		Instantaneous mixing efficiency\, $\eta$,
		}},
	columns/new col/.style={string type},
	/pgf/number format/precision=4,
	display columns/0/.style ={column name=},
	display columns/1/.style ={column name=Full plume},
	display columns/2/.style ={column name=Class U},
	display columns/3/.style ={column name=Class T},
	display columns/4/.style ={column name=Class A},
	columns={new col, 0,1,2,3}
	]{data.csv}\\
	\caption{Volume-averaged mixing quantities $\varepsilon, \chi, I, \partial_z b$ and mixing efficiency $\eta$ 
 defined in \S\ref{sec:mixvars} at $t=15$. Averages are computed over the full plume and within class U, T and 
 A. The percentage of the full plume volume associated with each class is given to indicate the relative 
 contribution of each class to the full plume average.}
	\label{tab:mixing}
	\end{center}
\end{table}

We summarise the mixing regimes described by the mixing diagnostics as follows. In class U, we find active turbulence 
with large dissipation of TKE, consistent with the undiluted plume being unaffected by the surrounding stratified 
environment due to the shielding effect. Since the undiluted plume fluid becomes well-mixed during its rise through 
the uniform layer, there are relatively small buoyancy gradients. As a result, there is relatively little PE 
dissipation. In class T, rising plume fluid impinges upon the 
more buoyant surrounding environmental fluid, generating particularly strong vertical buoyancy gradients. Horizontal 
buoyancy gradients are also generated by overturning. Turbulence advected upwards in class U is carried over into 
class T but gradually suppressed by the intense buoyancy gradients present in the plume cap, thus reducing 
the activity parameter. As the turbulent motion stirs these buoyancy gradients, significant PE dissipation 
occurs and consequently a large mixing efficiency is achieved, around $50\%$. Finally in class A, which 
eventually dominates the volume of the full plume as the intrusion grows, the interior of the intrusion becomes
well-mixed resulting in weak buoyancy gradients. There is some indication of a weak stratification and layering
forming within the intrusion. Secondary mixing processes between the intrusion and environmental fluid at the bottom 
of the stratified layer can introduce larger vertical buoyancy gradients and result in some PE dissipation. Overall, 
TKE dissipation is weak as the buoyant forces driving turbulent motion are weakened by earlier mixing. Whilst both TKE 
and PE dissipation are weak in class A, they are of similar magnitude and hence a greater mixing efficiency is 
achieved compared with class U. This could be attributed to the continued entrainment of environmental fluid above and 
below the intrusion, introducing small-scale buoyancy gradients which are acted upon by the weak turbulent motion.

%%%%%%%%%%%%%%%%%%%%%%%%%%%%%%%%%%%%%%%%%%%%%%%%%%%%%%%%%%%%%%%%%%%%%%%%%%%%%%%%%%%%%%%%%%%%%%%%%%%%%%%%%%%%%

\section{Conclusions}

In this paper we have analysed a large eddy simulation of a buoyant pure plume penetrating into a linearly stably 
stratified layer. We have outlined the buoyancy-tracer volume distribution formalism to examine tracer 
transport via turbulent mixing. Using this formalism, we developed a method for objectively partitioning 
buoyancy-tracer space into three regions based on the net change in volume due to mixing. Each of these regions 
identifies corresponds with a class of fluid lying in coherent regions of the plume in physical space. The method 
distinguishes undiluted plume fluid (class U) from mixtures of plume and environmental fluid. Mixed fluid is 
further partitioned into newly-generated mixtures in the plume cap that are actively mixing with the environment 
(class T), and fluid in the radially spreading intrusion that has already undergone significant mixing (class A). In 
buoyancy-tracer space, the intrusion corresponds with an accumulation region (corresponding with class A) where volume 
collects and homogenises. Active mixing with the environment in the plume cap moves volume from the source region, 
through a transport region (corresponding with class T), into the accumulation region. The accumulation region 
represents the majority of the plume volume at late times. To quantify the mixing regime in each class of fluid, we 
use the buoyancy variance dissipation rate, turbulent kinetic energy dissipation rate, vertical buoyancy gradient and 
an activity parameter as diagnostic variables in each sub-volume of the plume. The instantaneous mixing efficiency is 
also calculated by treating the buoyancy variance dissipation rate as a proxy for the energy dissipated in turbulent 
mixing that actually results in mixing of buoyancy. 

Our results demonstrate a three-stage mixing process in quasi-steady state penetration of a plume into a
stably stratified layer. In the first `plume mixing' stage (class U), turbulence within the undiluted rising plume 
homogenises the buoyancy-tracer distribution as fluid rises into the stratified layer. The turbulent motion near the 
centreline of the plume is relatively unaffected by the surrounding stratification owing to a shielding effect from 
the plume edge and surrounding intrusion. This homogenisation process sets the range of buoyancy and tracer 
concentration which is first exposed to the environment when fluid overturns near the maximum penetration height. The 
`primary mixing' stage occurs as rising fluid impinges on the more buoyant environment, establishing intense buoyancy 
gradients in the plume cap (class T). The mixing of undiluted plume fluid with the surrounding environment near 
$\zmax$ has a particularly large mixing efficiency. The maximum penetration height approximately determines the 
buoyancy of the environmental fluid involved in the primary mixing stage and sets the extent of the buoyancy-tracer 
convex envelope for the remainder of the mixing process. As newly-generated mixed fluid joins the intrusion and 
homogenises with fluid already in the intrusion (class A), the intensity of turbulence decreases and buoyancy 
gradients weaken. Several secondary mixing processes occur in the intrusion. This includes the entrainment of 
environmental fluid surrounding the intrusion, and mixing with small amounts of undiluted plume fluid at the edge of 
the rising plume that immediately join the intrusion without entering the plume cap. The volume of environmental 
fluid entrained into the intrusion during quasi-steady state dominates the volume into the plume as a whole at late 
times, but entrainment in the plume cap is `strongest' in the sense of the largest fractional rate of increase in volume.

The statistical properties of turbulence are different in each of the three stages. The undiluted plume core is the 
most turbulent (as measured by the activity parameter) with TKE dissipation significantly larger than PE 
dissipation. In the plume cap, the intense buoyancy gradients result in large PE dissipation and small TKE 
dissipation and hence the entrainment of the surrounding environment achieves a large mixing efficiency, with around 
50\% of the total energy dissipated by turbulence resulting in mixing. As mixed fluid homogenises in the intrusion and 
further environmental fluid is entrained, weak buoyancy gradients are continually introduced and eroded by weak 
turbulence with low TKE dissipation. The mixing efficiency in the intrusion is moderately large, though smaller than in the 
plume cap. 

Models of convective penetration which cannot resolve the processes responsible for mixing and entrainment 
must parameterise the effects of mixing on the flow. The markedly different statistics in each class suggest that each 
stage of mixing should be parameterised separately. Parameterisations of mixing in convective penetration could 
exploit the changing proportion of the full plume volume in each sub-region. For example, at early times before the 
formation of an intrusion, the plume is dominated by undiluted plume fluid in the plume core with intense turbulence 
but the mixing efficiency is small. At late times as fluid accumulates in the intrusion, the plume volume is dominated 
by the intrusion with weaker buoyancy gradients and turbulence and a greater mixing efficiency. Therefore the 
turbulent statistics associated with the full plume volume must change over time.  

The partitioning method presented here, as well as the buoyancy-tracer volume distribution formalism as a whole,
offers a way to analyse mixing in numerical simulations of stratified flows. Physical arguments can be made
that restrict the regions of buoyancy-tracer space accessible via mixing and consideration of terms in the
volume distribution budget equation \eqref{eq:VDbudget} highlight the mixing processes that occur and the resulting
tracer transport. 

%%%%%%%%%%%%%%%%%%%%%%%%%%%%%%%%%%%%%%%%%%%%%%%%%%%%%%%%%%%%%%%%%%%%%%%%%%%%%%%%%%%%%%%%%%%%%%%%%%%%%%%%%%%%%

\backsection[Acknowledgements]{We thank the three anonymous referees for their thorough comments that significantly improved the quality of this manuscript. We also thank Yves Morel and Alison Ming for their insightful comments and suggestions during preparation of this manuscript.}

\backsection[Funding]{CWP acknowledges funding from EPSRC grant EP/T517847/1.}

%%%%%%%%%%%%%%%%%%%%%%%%%%%%%%%%%%%%%%%%%%%%%%%%%%%%%%%%%%%%%%%%%%%%%%%%%%%%%%%%%%%%%%%%%%%%%%%%%%%%%%%%%%%%%

\bibliographystyle{jfm}

\bibliography{biblio}

\appendix

\section{Plume generation method}\label{app:A}

In \S\ref{sec:sim} we introduce the numerical setup for investigating convective penetration of a buoyant plume into 
a stably stratified layer. The buoyant plume is generated in a small forcing region at the bottom of the domain with 
the plume centreline at the middle of the computational domain, $x=y=0$. We use a volumetric forcing method in which 
the vertical velocity, buoyancy and tracer concentration are relaxed towards prescribed profiles in the shallow 
forcing region indicated in figure~\ref{fig:schematic}. The prescribed profiles are chosen to be the
far-field solutions of the \citet{mtt} (henceforth MTT) axisymmetric plume equations for a pure plume with source 
radius $r_0$ and integral source buoyancy flux 
$F_0 = 2\int_0^\infty \left. \overline{w}\overline{b}\right|_{z=-H}\,r\mathrm{d}r$. The plume 
carries a passive tracer with a source tracer flux that has a dimensional value identical to the source buoyancy flux, 
i.e. in non-dimensional terms $F_0^{(\phi)} = 2\int_0^\infty \left. \overline{w}\overline{\phi}\right|_{z=-H}\,r\mathrm{d}r = 
\mathcal{B}F_0$ where $\mathcal{B} = LT^{-2}$ and this excludes the normalisation of $\phi$ by its source value on the 
plume centreline which is performed in post-processing. The term `pure' plume refers to the relationship between the 
source fluxes and is quantified in terms of the flux-balance parameter $\Gamma$ introduced by \citep{morton1959} and 
defined as
\begin{equation}
    \Gamma = \frac{5F_b Q^2}{8\alpha M^{5/2}},
\end{equation}
where $\alpha$ is the entrainment coefficient and the integral volume flux $Q$, specific momentum flux $M$ and 
buoyancy flux $F_b$ are defined as
\begin{equation}
     Q = \int_0^\infty \overline{w}\, r \mathrm{d}r, \quad M = \int_0^\infty \overline{w}^2 \, r \mathrm{d}r, \quad F_b = \int_0^\infty \overline{w}\overline{b} \, r \mathrm{d}r, \label{eq:fluxesA}
\end{equation}
where $\overline{\cdot}$ denotes an azimuthal and time average. Note that \emph{in appendix~\ref{app:A} and~\ref{app:B} only} 
we use $M, F_b$ (and $F_\phi$ in appendix~\ref{app:B}) 
to refer to these integral fluxes, consistent with the notation used in the literature (e.g. \citet{hunt2005, 
mvr2016}). A forced plume, in which there is an excess of momentum relative to the buoyancy forcing, has $0 < \Gamma < 
1$ and the special case of a pure plume in which the momentum and buoyancy fluxes are balanced has $\Gamma = 1$. The 
pure plume state is stable to perturbations in the amount of buoyancy supplied \citep{hunt2005}. The vertical profiles 
$r_m(z), w_m(z), b_m(z), \phi_m(z)$ for the radius, vertical velocity, buoyancy and tracer concentration in the setup 
we consider are
\begin{align}
    r_m(z) &= \frac{6}{5}\alpha (z+H-z_v), \label{eq:plume1} \\
    w_m(z) &= \frac{5}{6\alpha} \left( \frac{9}{10}\alpha F_0\right)^{1/3} (z+H-z_v))^{-1/3}, \label{eq:plume2}\\
    b_m(z) &= \frac{5F_0}{6\alpha}\left(\frac{9}{10}\alpha F_0\right)^{-1/3} (z+H-z_v)))^{-5/3}, \label{eq:plume3}\\
    \phi_m(z) &= \frac{5F_0^{(\phi)}}{6\alpha}\left(\frac{9}{10}\alpha F_0\right)^{-1/3} (z+H-z_v)))^{-5/3}, \label{eq:plume4}
\end{align}
where $z_v = -\frac{5}{6\alpha}r_0$ is the virtual origin (which ensures a source radius $r_0$) and $\alpha = 0.11$ 
is the entrainment coefficient. Since the source tracer flux is the same as the source buoyancy flux, and $b$ and 
$\phi$ evolve identically in the uniform layer up to a linear factor, the profile $\phi_m(z)$ used for the passive 
tracer is the same as the profile $b_m(z)$ used for the buoyancy except with $F_0$ replaced by $F_0^{(\phi)}$.

The full structure towards which the vertical velocity, buoyancy and tracer concentration are forced uses the vertical 
profiles \eqref{eq:plume1}--\eqref{eq:plume4} with Gaussian radial profiles of width $r_m(z)$. Gaussian profiles have 
been shown to approximate experimental data well \citep{list1982,papanicolaou1988,shabbir1994}. Tests were carried out 
with various other radial profiles at the source, all of which result in a Gaussian profile in $w, b$ and $\phi$ far 
from the source (but below the stratified layer) where the plume is fully developed. The forcing on $w$, $b$ and 
$\phi$ is then
\begin{align}
    f_w(\boldsymbol{x},t) &= \frac{1}{\tau}\left[w(\boldsymbol{x},t) - 2w_m(z)\exp\left[-2\frac{x^2+y^2}       
    {r_m(z)^2}\right]\left(1+\frac{1}{10}\xi(t)\right)\right] f_m(z), \label{eq:wforcing}\\
    f_b(\boldsymbol{x},t) &= \frac{1}{\tau}\left[b(\boldsymbol{x},t) - 2b_m(z)\exp\left[-2\frac{x^2+y^2}
    {r_m(z)^2}\right]\left(1+\frac{1}{10}\xi(t)\right)\right] f_m(z), \label{eq:bforcing}\\
    f_\phi(\boldsymbol{x},t) &= \frac{1}{\tau}\left[\phi(\boldsymbol{x},t) - 2\phi_m(z)\exp\left[-2\frac{x^2+y^2}
    {r_m(z)^2}\right]\left(1+\frac{1}{10}\xi(t)\right)\right] f_m(z), \label{eq:phiforcing}
\end{align}
where $\xi(t)$ is a random number between -1 and 1, used to apply uncorrelated 10\% perturbations to the prescribed 
profiles at each step (note that the same perturbation is \emph{not} used for all profiles), to initiate turbulence. 
The factor $1/\tau$ controls the coupling strength with the momentum equations. The size of $\tau$ is arbitrary other 
than being small enough to control against dynamical variation and large enough to avoid numerical instability. The 
function $f_m(z)$ constrains the forcing to a thin layer at the base of the domain. We use
\begin{equation}
    f_m(z) = \frac{1}{2}\left(1 - \tanh \left( \frac{z+H-L_c}{L_p}\right)\right),
\end{equation}
where $L_c$ is the depth of the forcing region above $z=-H$ and $L_p$ controls how sharply the forcing decays 
above $z=-H+L_c$. As illustrated in figure~\ref{fig:schematic}, $f_m(z) \approx 1$ for $z \lessapprox -H+L_c$ 
and $f_m(z) \approx 0$ for $z \gtrapprox -H+L_c$. Whilst the forcing is applied across the entire domain, $f_m(z)$ 
limits the depth in which the forcing is non-zero and the exponential factor in \eqref{eq:wforcing}, 
\eqref{eq:bforcing} and \eqref{eq:phiforcing} constrains the forcing to small radii $x^2 + y^2 \lessapprox 
r_m(z)^2$. An additional perturbation is applied to each velocity component in the two grid layers above $z=-H+L_c$ 
to initiate turbulence, which develops as the plume rises through the uniform layer. We ensure that the plume has 
reached self-similarity (i.e. the turbulence and plume structure are fully developed) before penetrating the 
stratified layer -- see figure~\ref{fig:profiles} in appendix~\ref{app:B}.

The forcing method detailed here is non-standard. We found the typical method of generating a buoyant plume with a 
simple buoyancy gradient on the bottom boundary (e.g.\ \citet{mvr2016, pham2007}) to be unsuitable owing to pinching 
of the plume radius close to the bottom boundary where inflow dominates the diffusive boundary buoyancy flux. 
Pinching reduces control of the source radius and results in excessive numerical artefacts due to the horizontal 
pseudo-spectral method. Validation of the forcing method detailed here and the numerical scheme detailed in \S\ref{sec:sim} is discussed in appendix~\ref{app:B}.

\section{Validating the numerical method}\label{app:B}

\subsection{Resolution sensitivity test}
\begin{figure}
    \centering
    \includegraphics[width=\textwidth]{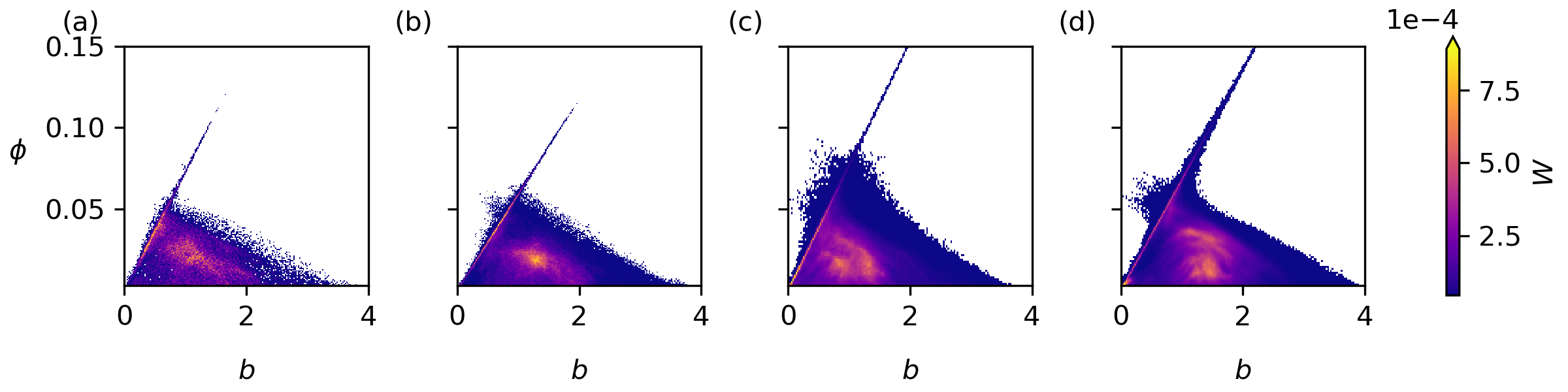}
    \caption{Buoyancy-tracer volume distribution $W(b, \phi; t)$, detailed in \S4, at $t=10$ in three simulations with resolution $128^2\times129, \,256^2 \times 257\, 512^2 \times 513$ and $1024^2 \times 1025$ left to right.}
    \label{fig:resolution}
\end{figure}

We study the problem presented here using large eddy simulations. Given the sub-grid scales are parameterised in LES,
it is necessary to validate the model to ensure that the quantities of interest are sufficiently well-resolved and 
that the results are not strongly dependent on the model resolution. Here, our primary focus is on the volume 
distribution in buoyancy/tracer space. Figure~\ref{fig:resolution} shows the volume distribution $W(b,\phi;t)$ at 
non-dimensional time $t=10$ for the simulations discussed in \S\ref{sec:sim} at three resolutions, $128^2\times129, 
\,256^2 \times 257$, \, $512^2 \times 513$, and $1024^2 \times 1025$, with all other aspects of the simulation setup 
fixed. To aid comparison, $W$ is normalised by the full plume volume $\sum_{ij} W_{ij}(t)$ (see \S\ref{sec:defn}). The 
structure of the distribution is similar at all resolutions. There is some noise in the distribution at lower 
resolution, since the smaller number of grid cells offers a smaller sample of the values of buoyancy and tracer 
concentration. The only element of the volume distribution structure that notably changes with resolution is the 
extent of the source line where $b \propto \phi$. This can be attributed to poor representation of the forcing profile 
at lower resolution; in the forcing region 
(see figure~\ref{fig:schematic}) the plume is thin compared with the width of the domain. This means relatively few 
grid points cover the plume at lower resolution and hence the forcing profile is poorly captured, in particular the 
maximum values of $b, w$ and $\phi$ on the plume centreline are reduced. Therefore, the forcing profile that 
\emph{is} achieved often has a smaller associated buoyancy and tracer source flux than the value $F_0$ prescribed for 
the simulation and hence the extent of the source line in buoyancy-tracer space is reduced. This is easily accounted 
for by increasing the prescribed value or decreasing the forcing relaxation timescale $\tau$ (see 
appendix~\ref{app:A}). In section~\ref{sec:flow} we report the value of $F_0$ computed from the 
simulation which suitably represents the plume forcing profiles achieved (see appendix~\ref{app:A} for a detailed 
discussion of the forcing method). 

\subsection{Numerical artefacts}
In the horizontal pseudo-spectral numerical scheme used by DIABLO, the 
spectral cutoff filter introduces unphysical oscillations in the scalar fields throughout a horizontal level in 
response to sharp gradients on the same horizontal level. This is known as Gibbs ringing (e.g. \citet{debonis2019}) 
and is qualitatively similar to the oscillations that arise in the Fourier series representation of a finite step 
function using a finite number of modes. In the simulation we present, these oscillations are relevant in two 
locations where particularly strong gradients of 
$b$ and $\phi$ arise. Firstly, at the top of the uniform layer before the plume penetrates the stratified layer. Here, 
$b$ and $\phi$ remain large on the plume centreline (and confined to small radii) whilst they vanish in the 
surrounding environment, resulting in steep gradients in both scalar fields. Note this ringing is present throughout 
the uniform layer and can be observed in figure~\ref{fig:evol} 
at the bottom of the domain in particular, but does not have a significant impact on the plume \emph{on average} (see 
azimuthally averaged profiles in figure~\ref{fig:profiles}) nor on the simulated buoyancy and tracer fluxes (see 
figure~\ref{fig:QMF}). Secondly, above the intrusion the plume width is small compared to the size of the domain and 
the plume buoyancy is significantly smaller than the surrounding environment, creating steep gradients in the 
buoyancy field in particular. The resulting oscillations imprint on the value of the buoyancy in the plume, shifting 
a small amount of volume away from the source line in buoyancy-tracer space and producing a `bulge' in the volume 
distribution $W$. This is evident in figure~\ref{fig:resolution}, particularly panels (c) and (d). Whilst the effect 
appears prominent (owing to the colourmap), this bulge contains only $0.77\%$ of the full plume volume at $t=10$ in 
the $512^2\times513$ simulation presented in the main paper. The numerical scheme conserves total tracer concentration 
in the absence of sources and sinks, though difficulties arise in calculating the plume volume that enters the 
stratified layer, since grid cells with $\phi < \phi_{\min}$ are excluded from the calculation of the source 
distribution $S$ but if the oscillation reverses sign then this tracer may be considered part of the plume. Similarly, 
the entrained volume $E(t)$ discussed in \S\ref{sec:entrainment} is underestimated due to these grid cells being 
excluded from the calculation of $F_{\phi}$. To account for Gibbs ringing, we therefore use the absolute value of 
the buoyancy and tracer concentration fields, $\lvert b \rvert$ and $\lvert \phi \rvert$, to calculate the source 
distribution $S$ and the mixing flux distribution $\boldsymbol{F}$. To aid clarity, 
we do not show the `bulge' in the volume 
distribution $W$ or the net mixing effect distribution $M$ in figures~\ref{fig:VD},~\ref{fig:Mplot} 
and~\ref{fig:M_partitioned}, but include the volume as part of class U when computing the volume of undiluted plume 
fluid $V(\mathcal{U})$. The regions in physical space corresponding to the erroneous volume are coloured according to 
the corresponding value of $b$ lying on the source line with the same value of $\phi$, i.e. we treat these values of 
$b$ and $\phi$ as part of class U and colour them accordingly when showing $M$ in physical space 
(see figure~\ref{fig:Mplot}).

\subsection{Plume representation}
\begin{figure}
    \centering
    \includegraphics[width=\textwidth]{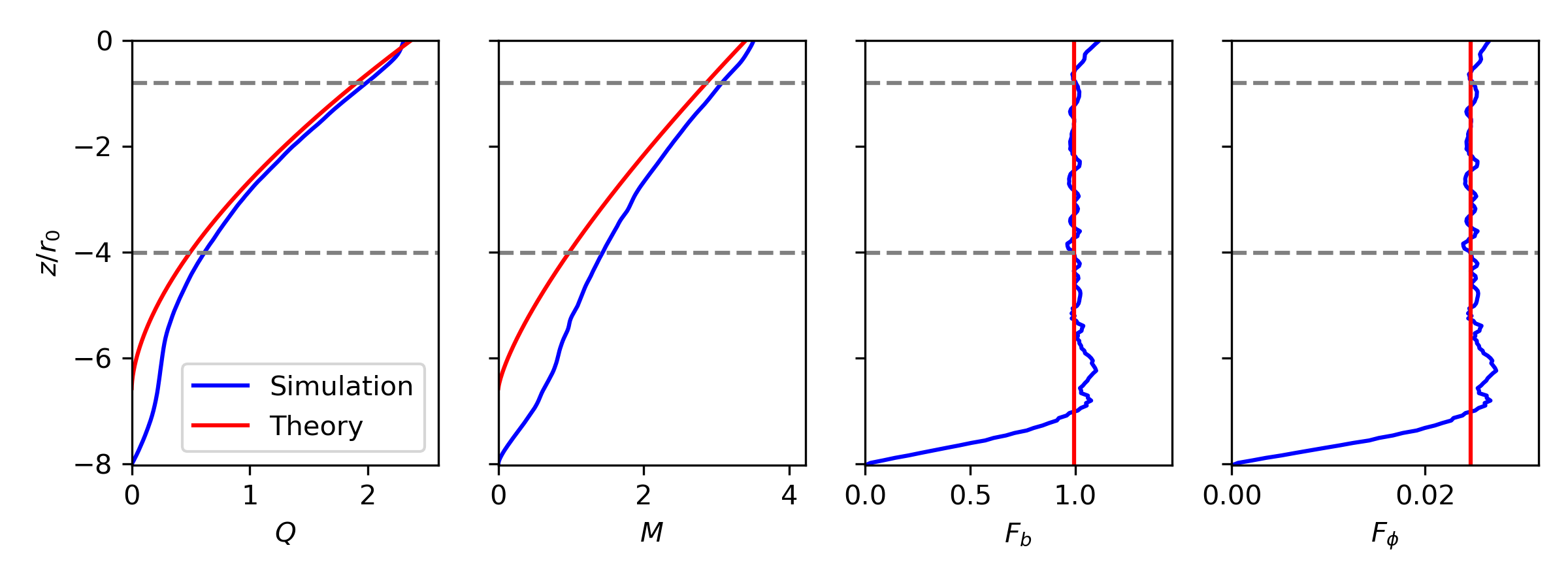}
    \caption{Integral volume flux $Q$, specific momentum flux $M$, buoyancy flux $F_b$ and tracer flux $F_\phi$
    defined in \eqref{eq:fluxesA} and \eqref{eq:fluxes} shown in the uniform layer $z \le 0$. Simulation data shown in 
    blue and theoretical predictions for a pure plume (using vertical profiles given in 
    \eqref{eq:plume1}--\eqref{eq:plume3} with a Gaussian radial profile) shown in red. Range of heights for which radial profiles 
    are shown in figure~\ref{fig:profiles} indicated by grey dashed lines.}
    \label{fig:QMF}
\end{figure}

We further validate our numerical method by comparing the simulated plume with the canonical plume theory of 
\citet{mtt} and direct numerical simulations of plumes in the literature, which themselves have 
been extensively compared with experimental studies. The integral plume theory of MTT predicts that in a buoyant 
plume rising from a point source (or equivalently a source with radius $r_0$ with a virtual origin at $-z_v$), 
the radius, vertical velocity, buoyancy and tracer concentration become self-similar with respect to 
characteristic scales 
\begin{equation}
    r_m = \frac{Q}{M^{1/2}}, \quad w_m = \frac{M}{Q}, \quad b_m = \frac{F_b}{Q}, \quad \phi_m = \frac{F_\phi}{Q}, 
    \label{eq:plumescales}
\end{equation}
where the integral volume flux $Q$, specific momentum flux $M$ and buoyancy flux $F_b$ are defined in \eqref{eq:fluxesA} and the tracer flux $F_\phi$ is defined as
\begin{equation}
    F_\phi = \int_0^\infty \overline{w}\overline{\phi}\, r \mathrm{d}r, \label{eq:fluxes}
\end{equation}
where $\overline{\cdot}$ denotes an azimuthal and time average. Note that \emph{in appendix~\ref{app:A} and~\ref{app:B} only} 
we use $M, F_b$ and $F_\phi$ to refer to these integral fluxes, consistent with the notation used in the literature 
(e.g. \citet{hunt2005, mvr2016}). The MTT plume equations may be solved to find the vertical 
profiles of the characteristic scales in \eqref{eq:plumescales} for a pure plume as given in 
\eqref{eq:plume1} -- \eqref{eq:plume3} and from these, vertical profiles for the integral fluxes can be computed. 
These theoretical predictions are compared with the simulation results in figure~\ref{fig:QMF} which demonstrates good 
agreement once turbulence in the plume becomes fully developed during rise through the uniform layer. 

\begin{figure}
    \centering
    \includegraphics[width=\textwidth]{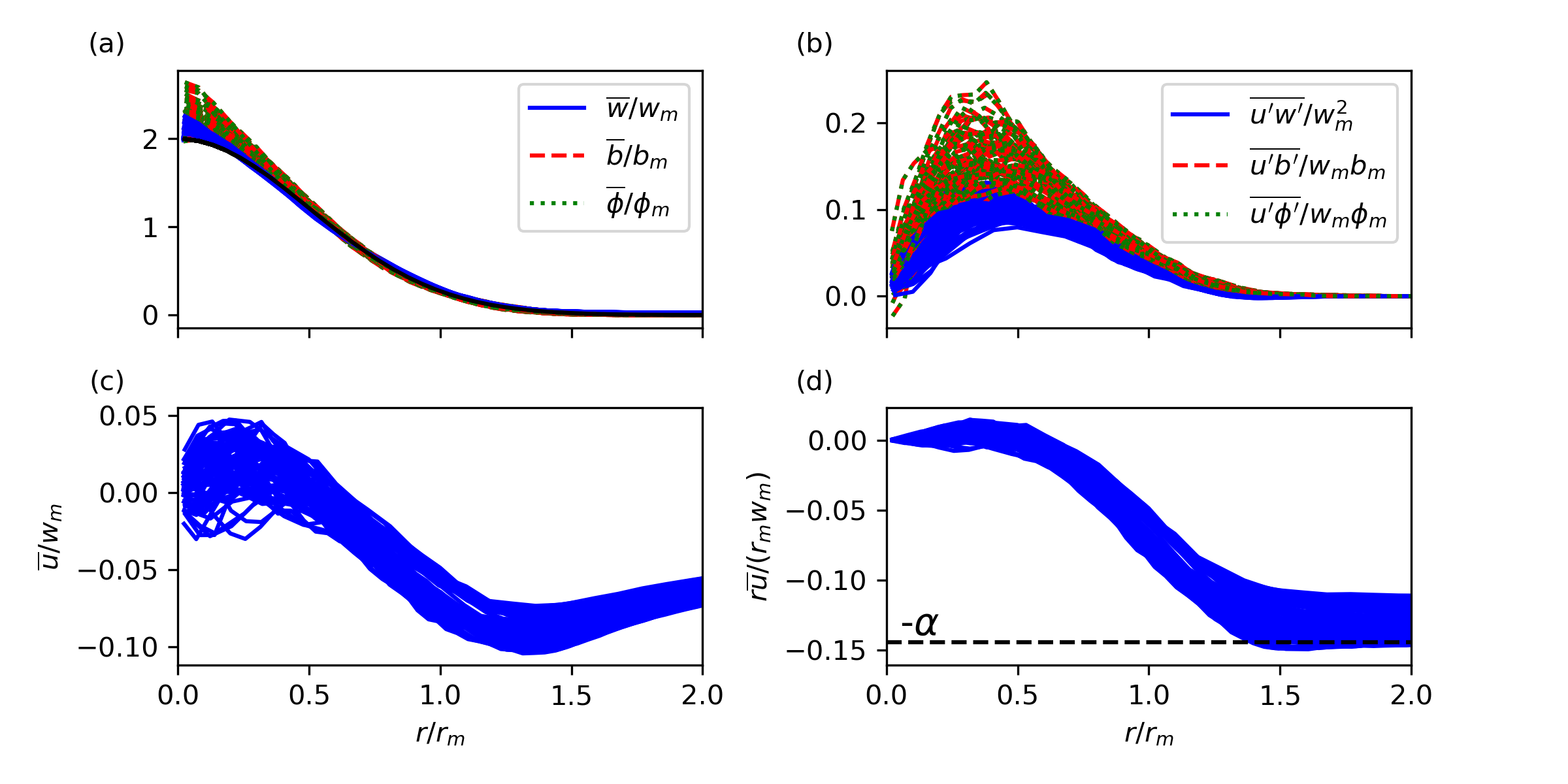}
    \caption{(a) Radial profiles of $\overline{w}, \overline{b}$ and $\overline{\phi}$. (b) Radial profiles of 
    radial momentum flux $\overline{u'w'}$, buoyancy flux $\overline{u'b'}$ and tracer flux 
    $\overline{u'\phi'}$. (c) Mean radial velocity $\overline{u}$. (d) Mean radial specific volume flux 
    $r\overline{u}$. All data shown in the interval $-5 < z < -1$ in the uniform layer, indicated by grey dashed 
    lines in figure~\ref{fig:QMF}. All profiles are self-similar with respect to the relevant scales defined in 
    \eqref{eq:plumescales}. In panel (d), the dashed line indicates the value of the entrainment coefficient
    $\alpha$ calculated from a linear fit of the plume radius scale $r_m$ and $z$.}
    \label{fig:profiles}
\end{figure}

In figure~\ref{fig:profiles}(a) we show vertical profiles of the time and azimuthally averaged vertical velocity 
$\overline{w}$, buoyancy $\overline{b}$, and tracer concentration $\overline{\phi}$, each normalised by the 
relevant characteristic scale. The profiles are from the region just below the stratified layer, indicated by the 
grey dashed lines in figure~\ref{fig:QMF}. The evident self-similarity of the plume, and the good convergence 
towards the experimentally observed radial Gaussian profile shown in black, demonstrate that the turbulent 
behaviour of the plume is well represented. This is further supported by the self-similarity of the 
radial momentum, buoyancy and tracer fluxes $\overline{u'w'}, \overline{u'b'}$, and $\overline{u'\phi'}$ when 
normalised by the relevant scales in figure~\ref{fig:profiles}(b). Here, $u$ refers to the \emph{radial} velocity 
and the prime notation refers to the turbulent component. The integral theory of MTT does not make predictions for 
these radial fluxes but they are in close agreement with the direct numerical simulations presented in 
\citet{mvr2016}. Figure~\ref{fig:profiles}(c) shows the mean radial velocity $\overline{u}$ which is also in 
agreement with DNS but with increased spread in the profiles owing to the periodic boundaries in our setup, which 
modifies the flow into the plume when compared with open boundaries typically used in numerical simulations of 
plumes, e.g. \citet{mvr2016}. Nonetheless we show that the entrainment is well captured by considering the mean 
radial specific volume flux $r\overline{u}$ in figure~\ref{fig:profiles}(d). The MTT entrainment hypothesis states 
$\left[ ru\right]_{r=\infty} = -\alpha r_m w_m$, where $\alpha$ is the entrainment coefficient, which relates the 
vertical velocity with the radial inflow. The value of $\alpha$ computed in our numerical simulations is shown in 
figure~\ref{fig:profiles}(d) which demonstrates the entrainment behaviour found in experimental observations of plumes.
The turbulent statistics in the plume are further validated by following the analysis presented in \citet{mvr2016}. 
For example, the precise characteristics of the turbulence are tested by calculating the invariants of the anisotropy 
tensor as detailed in \citet{lumley1977}. Our results indicate turbulence with weak anisotropy and axisymmetry (not 
shown), agreeing with the direct numerical simulations of \citet{mvr2016}.

\section{Buoyancy-tracer volume distribution evolution equation}\label{app:C}
In \S\ref{sec:defn} we define the volume distribution $W(B, \Phi; t)$ and its governing equation 
\begin{equation}
	\frac{\partial W}{\partial t} = - \bnabla_{(B, \Phi)} \cdot \boldsymbol{F} + S, \label{eq:VDbudgetAPP}
\end{equation}
where $\boldsymbol{F}$ and $S$ represent the flux and boundary source/sink of $W$ respectively. We 
derive this evolution equation for the volume distribution $W$ by starting with two scalar fields 
$b(\boldsymbol{x},t)$ and $\phi(\boldsymbol{x},t)$ satisfying
\begin{align}
	\frac{\partial b}{\partial t} + \boldsymbol{u} \cdot \bnabla b &= \dot{b}, \label{eq:Bevol}\\
	\frac{\partial \phi}{\partial t} + \boldsymbol{u} \cdot \bnabla \phi &= \dot{\phi}, \label{eq:Phievol} 
\end{align}
with $\bnabla \cdot \boldsymbol{u} = 0$. Here, we use $\dot{b}$ and $\dot{\phi}$ to represent general forcing 
terms which are replaced in \S\ref{sec:defn} with the non-advective forcing on buoyancy $b$ and tracer concentration 
$\phi$. Consider a fixed volume $V$ in which \eqref{eq:Bevol} and \eqref{eq:Phievol} hold and define
\begin{equation}
	g_\mathcal{F} = \int_V \mathcal{F}(b, \phi) \mathrm{d}V,
\end{equation}
where $\mathcal{F}(b, \phi)$ is an arbitrary function of $b$ and $\phi$. Then
\begin{equation}
	\begin{aligned}
		\frac{\partial g_\mathcal{F}}{\partial t} 
			&= \int_V \frac{\partial \mathcal{F}}{\partial b}\frac{\partial b}{\partial t} + \frac{\partial
			\mathcal{F}}{\partial \phi}\frac{\partial \phi}{\partial t} \, \mathrm{d}V \\
			&= \int_V \frac{\partial \mathcal{F}}{\partial b} \left( \dot{b} - \boldsymbol{u} \cdot \bnabla b
			\right) + \frac{\partial \mathcal{F}}{\partial \phi} \left( \dot{\phi} - \boldsymbol{u} \cdot \bnabla \phi
			\right) \, \mathrm{d}V \\
			&= \int_V \frac{\partial \mathcal{F}}{\partial b}\dot{b} + \frac{\partial \mathcal{F}}{\partial \phi}\dot{\phi}
			\, \mathrm{d}V - \int_V \left(\boldsymbol{u}\cdot\bnabla b\right) \frac{\partial \mathcal{F}}{\partial
			b} + \left(\boldsymbol{u}\cdot\bnabla\phi\right) \frac{\partial \mathcal{F}}{\partial \phi} \,
			\mathrm{d}V \\
			&= \int_V \frac{\partial \mathcal{F}}{\partial b}\dot{b} + \frac{\partial \mathcal{F}}{\partial \phi}\dot{\phi}
			\, \mathrm{d}V - \int_V \bnabla \cdot \left( \boldsymbol{u} \mathcal{F}(b, \phi) \right) \, \mathrm{d}V,
	\end{aligned}
\end{equation}
since $\bnabla \cdot \boldsymbol{u} = 0$. By the divergence theorem we have
\begin{equation}
	\frac{\partial g_\mathcal{F}}{\partial t} = \int_V \frac{\partial \mathcal{F}}{\partial b}\dot{b} +
				\frac{\partial \mathcal{F}}{\partial \phi}\dot{\phi} \, \mathrm{d}V + 
				\int_{\partial V} \boldsymbol{u}\cdot\boldsymbol{n} \,\mathcal{F}(b, \phi) \, \mathrm{d}S,
	\label{eq:budget1}
\end{equation}
where $\partial V$ is the boundary of $V$ and $\boldsymbol{n}$ is the \emph{inward} normal on $\partial V$. This 
forms an evolution equation for $g_\mathcal{F}$. Note that we choose an inward normal so that the final term is 
positive when fluid flows \emph{into} $V$. 

We now consider the specific choice $\mathcal{F}(b,\phi; B, \Phi) = I(b; B) I(\phi; \Phi)$ where, for a field 
$\psi(\boldsymbol{x},t)$ defined in $V$, $I(\psi; \Psi)$ is the indicator function for the subset of $V$ where $
\psi(\boldsymbol{x},t) > \Psi$, i.e. 
\begin{equation}
    I(\psi; \Psi) = 
        \begin{cases}
            1 & \psi(\boldsymbol{x}, t) > \Psi\\
            0 & \psi(\boldsymbol{x}, t) \le \Psi.
        \end{cases}
\end{equation}
With this choice of $\mathcal{F}$, the function $g_\mathcal{F}$ is the volume of fluid in $V$ with $b > B$ and 
$\phi > \Phi$. Furthermore $\frac{\partial^2}{\partial B \partial \Phi} g_\mathcal{F}$ is the `volume density', 
i.e. $\frac{\partial^2}{\partial B\partial \Phi}g_\mathcal{F} \delta b \delta \phi$ is the volume of fluid in $V$ 
with $B < b(\boldsymbol{x},t) < B + \delta b$ and $\Phi < \phi(\boldsymbol{x},t) < \Phi + \delta \phi$. This 
leads to the choice of name `buoyancy-tracer volume distribution' for $W$.

We now simplify the right-hand side terms in \eqref{eq:budget1} for the choice $\mathcal{F}(b, \phi; B, \Phi) = 
I(b; B)I(\phi; \Phi)$. We have
\begin{equation}
	\begin{aligned}
		\int_V \frac{\partial \mathcal{F}}{\partial b} \dot{b} \, \mathrm{d}V 
				&= \int_V \dot{b} \,\delta(b(\boldsymbol{x},t) - B) I(\phi; \Phi) \, \mathrm{d}V \\
				&= \int_{S(B, \Phi)} \dot{b} \,\frac{\mathrm{d}S}{\partial b / \partial \boldsymbol{m}} \\
				&= -\frac{\partial}{\partial B} \int_V \dot{b} \, I(b; B) I(\phi; \Phi)
				\mathrm{d}V,
	\end{aligned}
\end{equation}
where $S(B, \Phi)$ is the surface in $V$ where $b(\boldsymbol{x},t) = B$ and $\phi(\boldsymbol{x},t) > \Phi$, 
and $\boldsymbol{m}$ is the normal to the surface $S(B, \Phi)$. Similarly,
\begin{equation}
	\int_V \frac{\partial \mathcal{F}}{\partial \phi} \dot{\phi} \, \mathrm{d}V = -\frac{\partial}{\partial 
        \Phi}\int_V \dot{\phi} I(b; B) I(\phi; \Phi) \, \mathrm{d}V.
\end{equation}
Then the evolution equation \eqref{eq:budget1} with the choice $\mathcal{F}(b, \phi; B, \Phi) = I(b; B) I(\phi; 
\Phi)$ gives the integral form \eqref{eq:VDcumbudget} of the evolution equation \eqref{eq:VDbudgetAPP}, which 
governs the buoyancy-tracer \emph{cumulative} volume distribution:
\begin{equation}
	\begin{aligned}
		\frac{\partial }{\partial t} \int_V I(b; B) I(\phi; \Phi) \, \mathrm{d}V
			&= -\frac{\partial}{\partial B} \left[ \int_V \dot{b} \,I(b; B) I(\phi;\Phi)\,
				\mathrm{d}V \right] \\
			&- \frac{\partial}{\partial \Phi}\left[ \int_V \dot{\phi} \,I(b; B) I(\phi;
				\Phi) \, \mathrm{d}V\right] \\
			&+ \int_{\partial V} \boldsymbol{u}\cdot\boldsymbol{n} \, I(b; B) I(\phi; \Phi) \,
			\mathrm{d}S.
	\end{aligned}\label{eq:VDcumbudget}
\end{equation}
The governing equation for the buoyancy-tracer volume distribution $W(B, \Phi; t)$ is obtained by taking 
$\partial^2/\partial B \partial \Phi$ of \eqref{eq:VDcumbudget} to get
\begin{equation}
	\begin{aligned}
		\frac{\partial}{\partial t} \int_V \delta(b(\boldsymbol{x},t) - B) \delta(\phi(\boldsymbol{x},t) - \Phi) \, \mathrm{d}V
			&= -\frac{\partial}{\partial B} \left[ \int_V \dot{b} \,\delta(b(\boldsymbol{x},t) - B) \delta(\phi(\boldsymbol{x},t)-\Phi)\,
				\mathrm{d}V \right] \\
			&- \frac{\partial}{\partial \Phi}\left[ \int_V \dot{\phi} \,\delta(b(\boldsymbol{x},t)- B) \delta(\phi(\boldsymbol{x},t)-
				\Phi) \, \mathrm{d}V\right] \\
			&+ \int_{\partial V} \boldsymbol{u}\cdot\boldsymbol{n} \, \delta(b(\boldsymbol{x},t)- B) \delta(\phi(\boldsymbol{x},t)- \Phi) \,
			\mathrm{d}S.\label{eq:VDbudgetAPP2}
	\end{aligned}
\end{equation}
Since each integral is taken over the volume $V$, we are left with functions of $B, \Phi$ and $t$ alone. With the 
definitions
\begin{align}
    W(B, \Phi; t) &= \int_V \delta(b(\boldsymbol{x},t) - B) \delta(\phi(\boldsymbol{x},t) - \Phi) \, \mathrm{d}V, \\
    F_b(B, \Phi; t) &= \int_V \dot{b} \,\delta(b(\boldsymbol{x},t) - B) \delta(\phi(\boldsymbol{x},t)-\Phi)\,
				\mathrm{d}V, \\
    F_\phi(B, \Phi; t) &= \int_V \dot{\phi} \,\delta(b(\boldsymbol{x},t)- B) \delta(\phi(\boldsymbol{x},t)-
				\Phi) \, \mathrm{d}V,\\
    S(B, \Phi; t) &= \int_{\partial V} \boldsymbol{u}\cdot\boldsymbol{n} \, \delta(b(\boldsymbol{x},t)- B) \delta(\phi(\boldsymbol{x},t)- \Phi) \,
			\mathrm{d}S,
\end{align}
then \eqref{eq:VDbudgetAPP2} can be written as
\begin{equation}
    \frac{\partial W}{\partial t} = - \bnabla_{(B, \Phi)} \cdot \boldsymbol{F} + S,
\end{equation}
where $\boldsymbol{F} = (F_b, F_\phi)$, which completes the derivation.

\section{Numerical method for identifying the accumulation region of $(b,\phi)$-space}
\label{app:D}

In \S\ref{sec:partitioning} we introduce a partitioning of buoyancy-tracer space into three regions. Undiluted plume 
fluid corresponds with regions of buoyancy-tracer space where $M(b, \phi; t) < 0$ in quasi-steady state (see 
\S\ref{sec:qss}). Mixed fluid in the stratified layer is identified by regions of $(b,\phi)$-space with $M(b, \phi; 
t) > 0$. Mixed fluid is further partitioned by a threshold value $m^*(t)$ such that $0 < M \le m^*$ identifies 
newly-generated mixed fluid, where plume fluid is actively mixing with the environment in the plume cap; and $M > m^*$ 
identifies plume fluid which has mixed with the environment and is accumulating	in the radially spreading intrusion.

The threshold $m^*(t)$ is chosen to minimise the total mixing flux convergence $f(m; t)$ of the transport region 
$\mathcal{T} =\{0 < M \le m^*\}$, i.e. $m^*(t)$ is the value $m$ that minimises $\left| f(m; t) \right|$ where
\begin{equation}
	f(m; t) = \int_{\{ 0 < M \le m \}} \frac{\mathrm{d} W}{\mathrm{d} t} - S \, \mathrm{d}b\mathrm{d}\phi =
	-\int_{\{0< M\le m\}} \bnabla_{(b,\phi)} \cdot \boldsymbol{F} \, \mathrm{d}b\mathrm{d}\phi.
	\label{eq:f}
\end{equation}
so that the volume that enters the region $\mathcal{T}$ is approximately equal to the volume leaving $\mathcal{T}$. 
The region $\mathcal{A}$ which represents the remainder of the mixed fluid must then capture the accumulation of mixed 
fluid.

The numerical implementation of this method involves, at each time step $k$, the following steps:
\begin{enumerate}
	\item identify the current maximum value of $M$ throughout buoyancy-tracer space, denoted $\mathcal{M} = 
            \max_{i, j} M_{ij}(t_k)$;
	\item for each of $N_M$ test values of $M$, $m = 0, \dots, \mathcal{M}$, calculate 
		\begin{equation}
			f(m; t_k) \approx \sum_{i, j \vert 0 < M_{ij} \le m} \left[ \frac{W_{ij}(t_{k+1}) - 
					W_{ij}(t_k)}{\Delta t} - S_{ij}(t_k) \right];
		\end{equation}
	\item identify the test value $\tilde{m}(t_k)$ which minimises $f(m; t_k)$.
\end{enumerate}
Then, once a threshold $\tilde{m}(t_k)$ has been chosen for each timestep $t_k$, the final threshold $m^*(t)$
is chosen by applying a rolling average with an appropriate number of timesteps. We choose $N_M = 200$ and a
rolling average width of 10 timesteps. The motivation for applying a rolling average is to smooth the
threshold $m^*(t)$ so that the regions $\mathcal{T}$ and $\mathcal{A}$ do not grow and shrink dramatically in
response to short-term changes in the flux divergence.

\begin{figure}
	\centering
	\includegraphics[width=\textwidth]{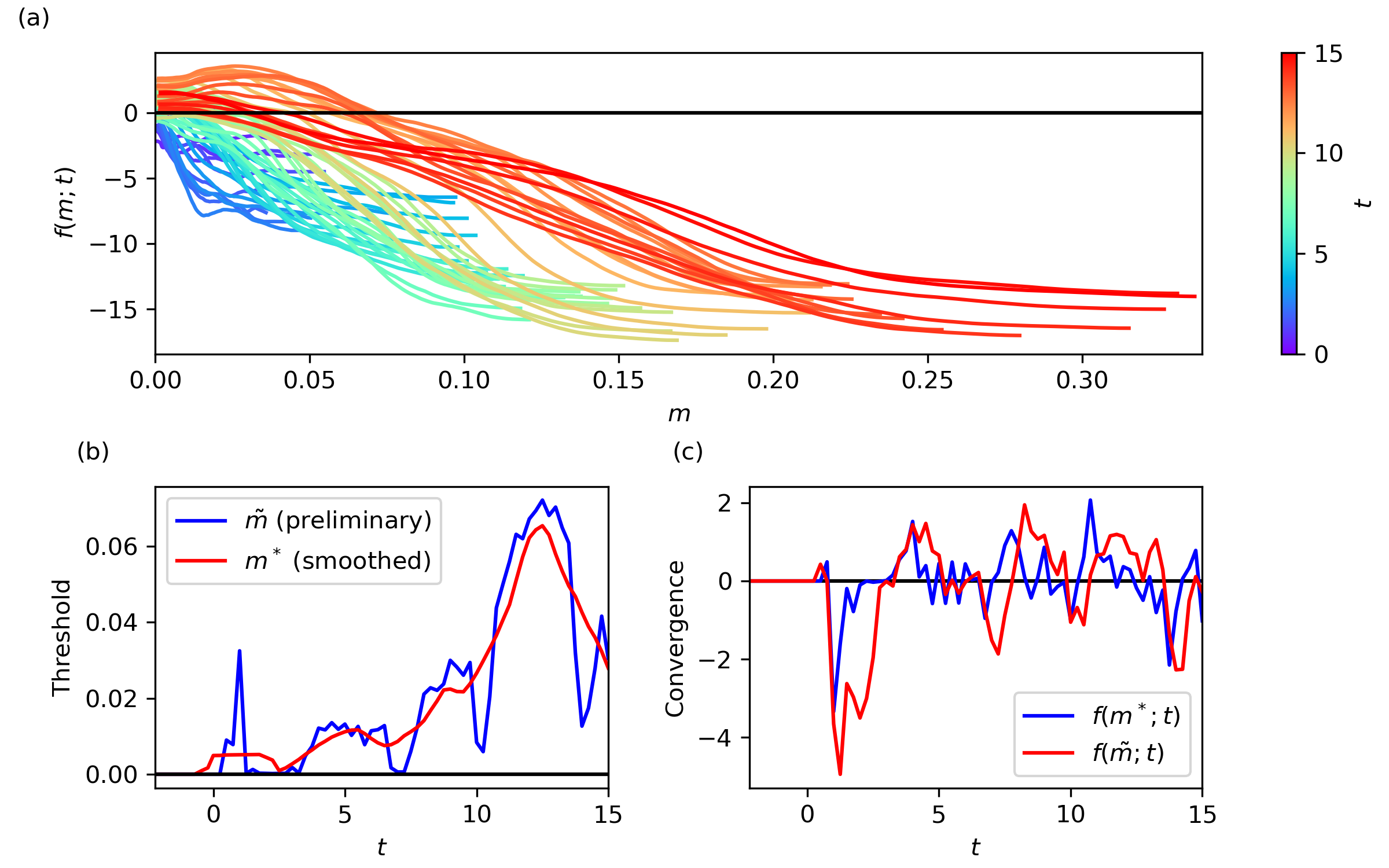}
	\caption{(a) Total flux divergence $f(m; t)$ defined in \eqref{eq:f} and the resulting preliminary and
		smoothed threshold choice in (b). Error in $f$ from smoothing process shown in (c).}
	\label{fig:m_thresh_calc}
\end{figure}

Figure~\ref{fig:m_thresh_calc}(a) shows the total flux convergence $f(m; t)$ for all times post-penetration and
figure~\ref{fig:m_thresh_calc}(b) shows the corresponding preliminary thresholds $\tilde{m}(t)$ and the final
thresholds $m^*(t)$. The total mixing flux convergence of class T for the preliminary and final thresholds over time
is shown in figure~\ref{fig:m_thresh_calc}(c), indicating the (generally small) error introduced by smoothing
the thresholds.

\section{Potential energy budget}
\label{app:E}
In section~\ref{sec:mixvars} we introduce the perturbation potential energy $E_p = \frac{1}{2}\langle b'^2 \rangle$ where 
$b'$ is the departure from the initial linear stratification, i.e. $b(\boldsymbol{x}, t) = b'(\boldsymbol{x},t) + z$. This 
form of the potential energy may be derived from equation (2.15) of \citet{holliday1981} under the assumption of a 
constant buoyancy gradient in the initial stratified environment. The perturbation PE may be treated as a proxy for 
available potential energy. To examine the mixing processes in the plume we wish to identify the primary sink of $E_p$ in 
order to calculate the mixing efficiency. We must therefore form a budget equation for the perturbation potential energy 
$E_p$.

We start with the governing equations \eqref{eq:ns1} -- \eqref{eq:ns3} including the SGS terms and drop the
hat notation. Substituting the buoyancy decomposition defined above, $b = z + b'$, the buoyancy evolution equation becomes
\begin{equation}
	\frac{\partial b'}{\partial t} + w + \boldsymbol{u\cdot}\bnabla b' = \frac{1}{\Rey\Pr}\bnabla^2 b' 
            + \nabla \cdot \left( \kappaSGS^{(b)} \nabla b' \right) + \frac{\partial}{\partial z} \kappaSGS^{(b)} 
            + f_b.
\end{equation}
Now, multiplying by the buoyancy departure from the initial stratification $b'$ and volume averaging over a fixed volume $V$ gives
\begin{equation}
	\frac{\mathrm{d} E_p}{\mathrm{d}t} = 
        \int_{\partial V} \left[ \frac{1}{\Rey \Pr} + \kappaSGS^{(b)}\right] b'\bnabla b'\cdot\boldsymbol{\mathrm{d}S} - \overline{\chi}
	- \overline{J_b} - \langle b' \frac{\partial\kappaSGS^{(b)}}{\partial z} \rangle
\end{equation}
where $\langle \cdot \rangle$ denotes a volume average over $V$. Note that the term involving $f_b$ has been 
neglected since the buoyancy forcing vanishes above the forcing region and we will apply this perturbation PE budget in 
the stratified layer. The first term on the LHS represents the diffusive buoyancy flux across the boundary $\partial 
V$, which is non-zero only where the plume penetrates the stratified layer. The second term is the volume averaged 
buoyancy variance dissipation rate
\begin{equation}
	\overline{\chi} = \left\langle \left(\frac{1}{\Rey\Pr} + \kappaSGS^{(b)}\right) \,|\bnabla b' |^2 \right\rangle, 
\end{equation}
which represents the primary sink of perturbation potential energy. The third term is the volume averaged buoyancy 
flux 
\begin{equation}
	\overline{J_b} = \langle b' w \rangle
\end{equation}
which represents an exchange between kinetic and potential energy. The last term captures the effect of the
spatially-varying SGS diffusivity acting on the background stratification. 

\end{document}